\documentclass[letterpaper]{article} 
\usepackage{aaai2026}  
\usepackage{times}  
\usepackage{helvet}  
\usepackage{courier}  
\usepackage[hyphens]{url}  
\usepackage{graphicx} 
\usepackage{natbib}  
\usepackage{caption} 
\usepackage{algorithm}
\usepackage{algorithmic}
\usepackage{amsmath}
\usepackage{booktabs}   
\usepackage{tabularx}   
\newcolumntype{Y}{>{\centering\arraybackslash}X} 
\usepackage{amsfonts}
\usepackage{amssymb}
\usepackage{pifont}
\newcommand{\cmark}{\ding{51} }%
\newcommand{\xmark}{\ding{55} }%
\newcommand{\pmark}{\ding{61} }%
\usepackage[most]{tcolorbox}
\tcbset{before skip=6pt, after skip=12pt}
\usepackage{tcolorbox}

\makeatletter
\@ifundefined{isChecklistMainFile}{
  \newif\ifreproStandalone
  \reproStandalonetrue
}{
  \newif\ifreproStandalone
  \reproStandalonefalse
}
\makeatother

\ifreproStandalone
\usepackage{aaai2026}
\usepackage{times}
\usepackage{helvet}
\usepackage{courier}
\usepackage{xcolor}
\usepackage{makecell}
\usepackage{newfloat}
\usepackage{listings}
\DeclareCaptionStyle{ruled}{labelfont=normalfont,labelsep=colon,strut=off} 
\floatstyle{ruled}
\newfloat{listing}{tb}{lst}{}
\floatname{listing}{Listing}
\usepackage{xcolor}

\newcommand{\name}{ShortageSim }
\newcommand{\bname}{\textbf{ShortageSim }}

\title{ShortageSim: Simulating Drug Shortages under Information Asymmetry}

\author{
    Mingxuan Cui\equalcontrib,
    Yilan Jiang\equalcontrib,
    Duo Zhou\equalcontrib,
    Cheng Qian,
    Yuji Zhang\thanks{Corresponding authors.},
    Qiong Wang\footnotemark[2]
}
\affiliations{
    University of Illinois Urbana-Champaign\\

    \{mc96, yilanj2, duozhou2, chengq9, yujiz, qwang04\}@illinois.edu
}

\begin{document}

\maketitle

\begin{abstract}
Drug shortages pose critical risks to patient care and healthcare systems worldwide, yet the effectiveness of regulatory interventions remains poorly understood due to information asymmetries in pharmaceutical supply chains. 
We propose \textbf{ShortageSim}, addresses this challenge by providing the first simulation framework that evaluates the impact of regulatory interventions on competition dynamics under information asymmetry.
Using Large Language Model (LLM)-based agents, the framework models the strategic decisions of drug manufacturers and institutional buyers, in response to shortage alerts given by the regulatory agency. 
Unlike traditional game theory models that assume perfect rationality and complete information, \name simulates heterogeneous interpretations on regulatory announcements and the resulting decisions. 
Experiments on self-processed dataset of historical shortage events show that \name reduces the resolution lag for production disruption cases by up to 84\%, achieving closer alignment to real-world trajectories than the zero-shot baseline.
Our framework confirms the effect of regulatory alert in addressing shortages and introduces a new method for understanding competition in multi-stage environments under uncertainty.
We open-source \name and a dataset of 2,925 FDA shortage events, providing a novel framework for future research on policy design and testing in supply chains under information asymmetry.
\begin{links}
\link{Code}{https://github.com/Lemutisme/ShortageSim}
\end{links}
\end{abstract}

\section{Introduction}
Every day, clinicians across the globe face an painful decision: which patient receives the last piece of a life-saving medication? Drug shortages have escalated from occasional supply disruptions to a persistent global crisis threatening the foundation of modern healthcare. Before the COVID-19 pandemic exposed these vulnerabilities to the public, the global pharmaceutical supply chain has already experienced multiple crises. Canada \cite{videau2019canada}, Finland \cite{sarnola2019finland}, and France \cite{benhabib2020french} all reported unprecedented shortage levels, while the United States has averaged over 130 new shortage cases annually for the past decade \cite{ASHP2025statistics}. More alarmingly, the average shortage duration has surged from 9 months in 2011 \cite{GAO2011duration} to 14 months in 2016 \cite{GAO2016duration}, with some critical medications remaining unavailable for over 8 years.
Patients face delayed surgeries, substitution with less effective alternatives, or no treatment at all. Beyond the immeasurable costs of compromised patient outcomes, these shortages also impose substantial financial burdens. U.S. hospitals alone spend at least \$359 million annually managing shortage related logistics \cite{vizient2019cost}.

While drug shortages stem from many causes, this work focuses on supply disruption driven shortages, a main trigger of drug shortages in the U.S. These shortages often result from manufacturing capacity reductions due to quality failures, production line breakdowns, or regulatory compliance issues \cite{NAP2022resilience}. For example, two major manufacturers of propofol recalled their products and halted production in 2009, and the sole remaining supplier could not scale to meet the demand, triggering a national shortage \cite{Wookcock2013example}.
The propofol case reveals information asymmetry as a major factor that aggravates the drug shortage problem. Manufacturers guard production data as trade secrets, regulators operate with incomplete market visibility, and buyers cannot distinguish temporary disruptions from permanent exits. This information asymmetry transforms manageable supply shocks into prolonged crises, as stakeholders make decisions in isolation that collectively amplify shortages.

Information asymmetry in the U.S. drug supply chain stems primarily from two causes.
First, there exists a cost of information sharing. Production capacity and quality control data are commercially sensitive, and disclosing them can expose competitive weaknesses or invite regulatory examinations. 
Second, there is a persistent lack of effective communication channels between manufacturers, buyers, and regulators. \citep{NAP2022resilience} stated manufacturers are reluctant to invest without accurate demand information and regulatory agencies are receives limited information from manufacturers either. Comparative analyses in \cite{tadrous2024differences} further highlight that, unlike Canada’s centralized shortage registry, U.S. reporting remains decentralized and voluntary, delaying signal transmission to healthcare systems and making similar policies ineffective. The resulting information asymmetry means the regulator only sees aggregate shortage levels instead of  individual capacity losses or buyer inventories.

Recognizing the need for national coordination, the U.S. Food and Drug Administration (FDA) expanded its regulatory authority in 2012 by mandating manufacturers to report potential shortage risks \cite{FDA2013mandate}. The agency now maintains a public shortage database and issues alerts to inform the market about current or expected shortages. However, these interventions carry unintended risk of triggering hoarding behaviors to build safety stock and further exacerbate shortages. Traditional analytical frameworks using game-theoretic models struggle to capture how real-world decision-makers interpret ambiguous regulatory language, update beliefs dynamically, and respond to sequential market signals. Moreover, the subjective and sometimes irrational stockpiling behavior is often ignored.
The emerging Large Language Model (LLM) provides a novel toolset for alleviating the aforementioned limitations~\cite{wei2024towards, ai2025resmoe, ninggraph4mm}. To fill the scientific research gap of evaluating regulatory interventions, this work presents \textbf{ShortageSim}, an LLM-powered multi-agent framework for analyzing the role of information sharing through public announcement in addressing drug shortage problems. The framework models FDA regulators, pharmaceutical manufacturers, and healthcare buyers as autonomous agents navigating information-constrained environments. Each stakeholder interprets FDA announcements through its own analysis, forms beliefs about competitors' actions, and makes decisions that ripple through the supply chain.

By creating a controlled experimental environment to compare reactive and proactive policies, we confirm that proactive policies lead to increased stockpiling behavior and worse effect in resolving shortages.
\name enables counterfactual analysis of various FDA communication strategies and serves as a testbed for future research on policy design and evaluation. 

Our contributions are:

\begin{itemize}
    \item We introduce the first LLM-based multi-agent simulation framework for drug shortage management under information asymmetry. Beyond traditional mathematical models, our agents interpret regulatory announcements, infer market states from partial information and make context-dependent decisions under uncertainty.
    
    \item We collect, process and publicly release a dataset on 2,925 FDA shortage reports and 51 resolved historical event trajectories, addressing the current reliance of drug shortage research on synthetic or proprietary data.
    
    \item Our framework achieves consistently closer alignment with historical shortage trajectories than the zero-shot baseline across four LLM providers, reducing resolution lag for production disruption cases by up to 84\%.

    \item We provide a flexible testbed for counterfactual policy evaluation, enabling controlled evaluation of regulatory strategies, communication designs, and agent decision protocols in information constrained environments.

\end{itemize}

\section{Related Works}

\paragraph{Drug Shortage Management.}

Drug shortages in the United States are well documented in white papers, government reports, and case studies that examine their causes, impacts, and scope \citep{Kweder2013shortage, FOX2014shortage, Yurukoglu2017reimburse, mclaughlin2013effects, phuong2019impacts, Patel2021price, FDA2011duration, Edwards2021reimagine}. Most of these studies are descriptive, and root causes of the problem are yet to be investigated rigorously through analytical or empirical methods. 

Operations management (OM) literature views drug markets through the lens of optimization and control~\citep{Jia2017contract, Tucker2020incentivize, swinney2024mitigating, Zhao2025reimburse}.  The resulting static models assume rational decisions and overlook dynamic behaviors and information asymmetries~\cite{zhang-etal-2023-vibe}, thereby obscuring sparse but critical knowledge and limiting the accuracy of models trained on such data~\cite{zhang2024knowledge, zhang-etal-2025-law}.
Empirical evidence on policy effectiveness is likewise limited: although \citep{Lee2021mandate} show that mandatory disruption reporting mitigates shortages, the underlying mechanisms are still unclear. We address these gaps by examining how regulatory announcements influence supply chain decisions and shortage outcomes.

\paragraph{Agents for Pharmaceutical Supply Chain.}

Multi-agent systems have been extensively applied to pharmaceutical supply chains, though few has applied it to study the impact of regulations on mitigating drug shortages. Traditional frameworks demonstrate manufacturer-distributor interactions using platforms like MATLAB~\cite{pourghahreman2018agent} and AnyLogic~\cite{bozdougan2023agent}, focusing on disruption response and risk management strategies~\cite{Jing2020multi}. Recent drug shortage prediction models achieve high accuracy using machine learning approaches on historical data~\cite{liu2021development,pall2023predicting,postma2024identifying}. However, these systems treat regulatory actions as external constraints rather than modeling FDA as an active strategic agent, and prediction systems remain disconnected from dynamic stakeholder response modeling.

The integration of Large Language Models into multi-agent systems represents a paradigm shift in complex system simulations~\cite{li2023large, wang2024survey,li2024agent, qian2025modelingagent,qian2025userbench,xu2025instructagent}. Microsoft's OptiGuide framework ~\cite{li2023large} pioneered LLM integration for supply chain optimization. Recent frameworks demonstrate remarkable scaling and strategic reasoning capabilities~\cite{li2023camel,hong2024metagpt, zhang2025atomic, DBLP:conf/aaai/SongMK24}. Game-theoretic applications show that multi-agent LLM architectures significantly outperform single LLMs in simulating strategic behavior~\cite{li2024more}, while economic simulations like EconAgent~\cite{li2023econagent} demonstrate the ability to produce realistic macro-level phenomena arising from micro-level interactions.


\begin{figure*}[t]
\centering
\includegraphics[width=0.98\textwidth]{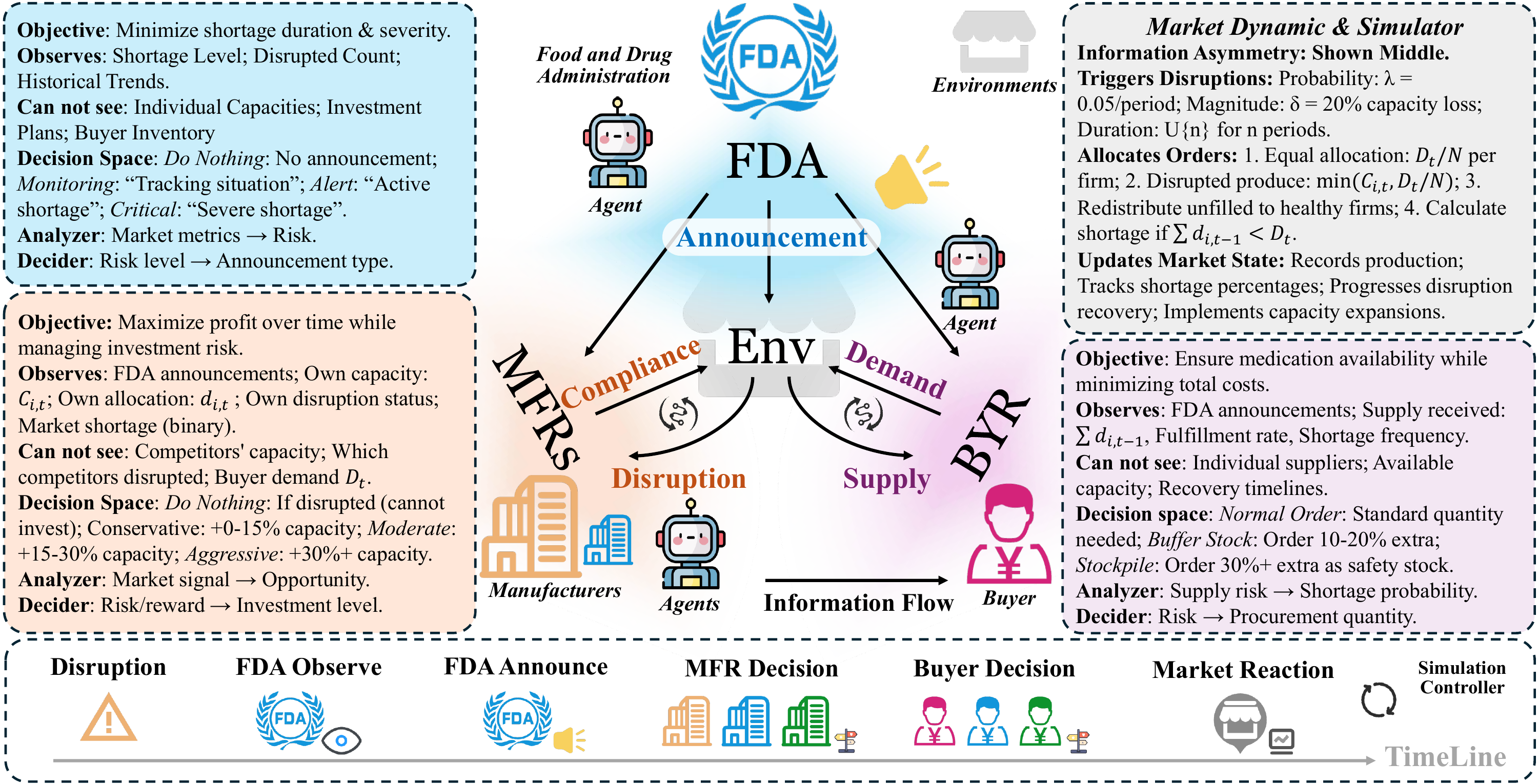}
\caption{\name Architecture. The framework models the drug supply chain dynamics through three agent roles: FDA regulator, buyer consortium, and $n$ competing manufacturers, coordinated by an Environment module. Each agent operates under information asymmetry (shown in \textbf{Information Availability}) and follows a two-stage LLM pipeline (Analyze $\rightarrow$ Decide). The timeline illustrates the sequential decision process from supply disruption through market reaction (details in \textbf{Sequential Decision Timeline}), capturing realistic stakeholder responses to regulatory announcements during drug shortage events.}
\label{fig_general}
\end{figure*}

\section{Methodology: Shortage Simulation Agents}
\subsection{System Architecture Overview}

To address the challenge of modeling decision-making under information asymmetry, \name employs a multi-agent simulation framework where LLM-powered agents interact within a drug supply chain. This framework design captures the nuanced interpretation of regulatory signals and market conditions, characterizing real-world responses to shortage events. \name models the market dynamics between manufacturers and buyers, under the regulatory interventions by the FDA during drug shortage events. All agents make decisions under partial information, reflecting the actual information asymmetry in drug supply chains.

As shown in Figure \ref{fig_general}, the framework consists of four core components: (1) the \textbf{Environment} module manages market dynamics and state transitions while maintaining information barriers among agents, (2) the \textbf{Agents} system implementing role-specific decision-making through LLMs that interpret partial information and form strategies, (3) the \textbf{Information Flow} that controls inter-agent communication to simulate the information asymmetries, and (4) the \textbf{Simulation Controller} that operates the overall execution flow and records all decisions for subsequent analysis.

\subsection{Market Mechanics and Dynamics}
We consider a drug market with $n$ competing \emph{manufacturers}, a consortium of health care providers as the \emph{buyer} and FDA as the \emph{regulator}. Each manufacturer $i$ possesses a \emph{production capacity} $c_i$. The drug in question is medically necessary, with \emph{patient demand} modeled deterministically as $D_0$. Initially, the market is assumed to be in equilibrium, meaning the total production exactly meets the patient demand $D_0$ until disruptions happen. 

We choose not to include intermediary entities like distributors and group purchasing organizations (GPOs) in the model, as they primarily facilitate logistics and negotiate prices respectively without affecting production capacity or demand dynamics. Their detailed operational data are not publicly available either. A detailed discussion of this modeling choice is provided in the Appendix.

\paragraph{Disruption Modeling.}

Our model accounts for internal supply disruptions on manufacturers' production capacities. At the start of each period, each manufacturer faces an independent probability $\lambda$ of experiencing disruption. When a disruption occurs to a manufacturer, its production capacity is reduced by a fixed fraction $\delta$ for $t$ periods and returns to the regular level afterwards, where $t$ is sampled from discrete uniform distributions.

\paragraph{Demand Allocation Mechanism.}
In the drug supply chain, buyers are observed to stockpile when they anticipate the cumulative orders will exceed the total production capacity \cite{sterman2015hoard}, driven by aversion to stockouts. To differentiate from the actual patient demand $D_0$, we define buyers' order quantity as \emph{demand}, denoted by $D$. Excess orders ($D - D_0$) are maintained as buyers' inventory for future periods.

In the nominal model, we assume all manufacturers are symmetric and the total demand $D$ is evenly allocated in equilibrium. If disruption affects a subset of manufacturers $M_D\subset[n] = \{1,2,\cdots,n\}$, their production becomes 
\begin{equation}\label{eq: disrupted q}
q_i = \min\left\{\frac{D}{n},\,c_i(1-\delta)\right\},\,i\in M_D.
\end{equation}
The resulting unfilled demand is 
$$
D_{\text{unfill}} = \sum_{i\in M_D}\big(D/n - c_i(1-\delta)\big)^+,
$$
where $x^+ = \max\{x,0\}$. This unfilled demand is reallocated evenly among other undisputed manufacturers $M_U=[n]\backslash M_D$, whose production is updated accordingly as  
\begin{equation}\label{eq: undisrupted q}
q_i = \min\left\{\frac Dn+\frac{D_{\text{unfill}}}{|M_U|},\,c_i\right\},\, i\in M_U.
\end{equation}
The total \emph{supply} is defined as $Q = \sum_{i\in[n]}q_i$ and \emph{shortage} is calculated as $(D-Q)^+$. We relax the symmetry assumption in the \textbf{Experiment} section to analyze different agent behavior under unequal market shares.

\paragraph{Financial Model}

The simulation's economic framework emphasizes the deviation between individual incentives and collective outcomes, highlighting why market coordination often fails during shortages. 

Each period undisputed manufacturers could ramp up their capacities at a unit \emph{cost} $Ct$ to fulfill the unmet patient demand. Capacity expansion typically involves constructing new production lines and is assumed to be available one time period after the investment decision, requiring manufacturers to anticipate future market conditions from regulatory signals. Moreover, if all manufacturers expand simultaneously based on the same signal, potential profits will be diluted by overcapacity, creating a coordination challenge.

On the other side, buyers pay the \emph{price} $p$ to purchase a unit of drug, incurring a \emph{holding cost} $h$ for unit leftover and a \emph{penalty} $s$ for unit unmet patient demand. As health providers, buyers are adverse to stockouts, but maintaining large safety stock imposes inventory costs and forces buyers to carefully calibrate their response to shortage signals. 

\subsection{Agent Design and Two-Stage Decision Pipeline}

\paragraph{Agent Architecture.}

Each agent in \name follows a two-stage decision pipeline of information processing followed by strategic decision-making. This architectural choice directly simulates the dynamics of decision making based on incomplete and ambiguous market signals among stakeholders. 
Firstly, \textbf{Collector \& Analyst} receives unstructured context, including FDA announcement text, aggregate market metrics, and historical demand trend, then produces a structured representation capturing the agent's interpretation of the current situation.
Secondly, \textbf{Decision Maker} takes this structured state representation and generates decisions with detailed reasoning. The two-stage design examines the effects of FDA communications through reasoning output and captures the discrepancies in the reasoning process across different agents.
More importantly, this stage does not require the output decisions to be optimal, an important assumption for realism. As \cite{croson2014coordination} observed, some stakeholders mistrust the rationality of their competitors' decisions and respond with suboptimal, even unreasonable choices.

\subsubsection{Role-Specific Agents Design}

\paragraph{Manufacturer Agents} represent pharmaceutical companies making capacity investment decisions under uncertainty about competitor status and future market conditions. The objective of each manufacturer is to maximize its own profit, by balancing the opportunity of profiting from competitors' disruptions and the financial losses from overcapacity when anticipated disruptions do not occur.
To support effective decision-making, manufacturer agents are instructed to: 1) understand the overall market structure, demand allocation mechanisms investment opportunity and risk of overcapacity; 2) infer market conditions from observable signals, regulatory announcements and the amount of demand it gets; 3) identify profit opportunities, assess market risks and anticipate competitors' potential decisions; 4) determine whether and how much to invest in capacity expansion.

\paragraph{Buyer Agent} represents a healthcare consortium facing a variant of the newsvendor problem under supply uncertainty. With patient safety as their top priority, the buyers can stockpile to build safety stock, but must balance the risk of stockout against the cost of purchasing and holding excess inventory.
To simulate realistic behavior, the buyer agent is guided to: 1) understand the market structure, own cost parameters and utility objectives; 2) interpret regulatory signal and detect changes in supply availability; 3) analyze current inventory levels and determine risk of stockout; 4) make purchasing decisions that balance the additional cost and the potential impact of stockouts.

\paragraph{FDA Agent} operates under a reactive policy framework, with a goal of mitigating drug shortage through public announcements without disclosing any individual stakeholder’s private information. The announcements should provide sufficient warnings to induce manufacturers' investment during shortage while preventing buyers' panic hoarding. To issue effective announcements, the FDA agent is instructed to: 1) understand market structure and its regulatory role in mitigating drug shortage; 2) detect supply-demand imbalances and predict potential shortages; 3) anticipate stakeholders' response to public announcements; and 4) determine the content and severity level of public announcements to balance urgency with stability.

\subsection{Information Asymmetry}

\paragraph{Sequential Decision Timeline.}
\label{sec:seq_dm}

As illustrated in Figure \ref{fig_general}, each simulation period follows an organized sequence of events that mimic information asymmetry in practice, ensuring agents make decisions based on partial information available to them in reality.
A period begins with a disruption phase in which affected firms learn their remaining capacities and recovery times. 
Next, after observing new disruptions and aggregate market outcomes from the previous period, the FDA decides whether to issue a public announcement. Unaffected manufacturers then simultaneously decide on capacity investments, based on allocated demand in the last period and the latest FDA announcements. 
The buyer simultaneously determines procurement quantities based on the amount of demand served in the last period and the same FDA announcements.
Finally, the market clears through the aforementioned allocation mechanism, with excess supply or shortage levels determined accordingly..

\paragraph{Information Availability.}
\label{sec:information_availability}
The framework enforces strict information separation that mirrors real pharmaceutical supply chains and complicates decision-making process.

\textbf{Manufacturers} possess complete knowledge of their own status, including current capacity, recovery time if being disrupted and past allocated demand. However, they cannot directly observe competitors status and must infer through regulatory signals, bringing challenges in building capacity investment strategies. For example, increase in its own demand is not sufficient for a manufacturer to determine whether the surge is caused by a disruption at another manufacturer or demand fluctuation on the buyer side.  

\textbf{Buyer} observes the total supply received each period, but not the allocation across manufacturers or their production capacities. Specifically, when a disruption happens, the buyer has no information on the remaining recovery time, bringing challenges to the buyer's purchasing strategy.

\textbf{FDA} operates outside the supply chain, and thus has no access to private information of manufacturers and the buyer except for aggregate shortage levels and the mandated report of disruptions.
Like the real-world agency, no information on capacity and investment of manufacturers and inventory levels of the buyer is visible to the FDA in our simulation .

Table~\ref{tab:info_availability} summarizes what variables are accessible to manufacturers (Mfrs), buyer, and the FDA, where \cmark indicates full availability, \xmark means no availability, and \pmark stands for partial availability, that a manufacturer can only observe its own status, but not those of their competitors.

\begin{table}[ht]
\centering
\begin{tabularx}{\linewidth}{lYYY}
\toprule
\textbf{Variable Privacy} & \textbf{Mfr.} & \textbf{Buyer} & \textbf{FDA} \\
\midrule
Disruption            & \pmark & \xmark & \cmark \\
Capacity              & \pmark & \xmark & \xmark \\
Investment            & \pmark & \xmark & \xmark \\
Allocated Demand      & \pmark & \xmark & \xmark \\
Total Supply          & \xmark & \cmark & \cmark \\
Total Demand          & \xmark & \cmark & \cmark \\
Shortage              & \xmark & \cmark & \cmark \\
Buyer Inventory       & \xmark & \cmark & \xmark \\
FDA Signal            & \cmark & \cmark & \cmark \\
\bottomrule
\end{tabularx}
\small
\caption{Information availability for each agent.  
\pmark = private/partial information; \cmark = known; \xmark = not known.}
\label{tab:info_availability}
\end{table}

\section{Experiments}
We evaluate \name on a self-processed FDA drug shortage dataset to verify its alignment with real-world trajectories and evaluate the effect of communication by FDA. The framework offers a new method for understanding competition in complex, multi-stage environments under uncertainty. It also provides a controlled experiment pipeline for systematically comparing regulatory policies and agent decision protocols in future research.


\subsection{Experiment Settings}

To evaluate the ability of \name reproducing historical shortage dynamics, we collect and preprocess 2,925 FDA reported drug shortage events and 51 resolved trajectories in 2023 and 2024. These cases are split into two ground-truth trajectory sets based on their reported shortage reasons. All reported drugs are grouped and identified by their National Drug Code (NDC), which uniquely links manufacturers to specific drug products. The \textbf{FDA-Disc} set comprises events that the FDA explicitly cites manufacturer production discontinuation as the primary cause, while the \textbf{FDA-NR} set contains cases with no specific cause reported. 
Each trajectory is truncated upon shortage resolution so its length represents the total resolution time in quarters. Data curation details are provided in Appendix.

Each simulation is specified by the number of manufacturers ($n$), shortage duration ($T$), disruption probability ($\lambda$), and disruption magnitude ($\delta$). One period corresponds to a calendar quarter. Following the FDA dataset statistics, we simulate cases with $T\in\{4,\dots,12\}$ and $n\in\{2,\dots,10\}$. Monopoly scenarios ($n=1$) are excluded because our model reduces to a trivial case with no decision to be made. For the FDA-Disc dataset, the disruption magnitude of each drug is estimated by the proportion of discontinued NDC packages reported by the FDA. Moreover, each trajectory starts with an enforced disruption and no separate disruption probability is specified. Details of parameter selection are described in the Appendix.
At each period~$t$, the framework records total demand, total supply, shortage amount, buyer's on-hand inventory, FDA announcements, and the set of disrupted manufacturers. These variables combined with all agent decisions generate a simulation trajectory.

 We test GPT-4o~\cite{hurst2024gpt}, Gemini 2.5 Flash~\cite{comanici2025gemini}, Claude Sonnet 4.5~\cite{Anthropic2025Claude4.5} and Deepseek V3.2 Exp~\cite{liu2024deepseek} with temperature $0.3$ as LLM backbones for all agents in our simulation. An ablation study on different temperatures is provided in the Appendix to show performance robustness. For comparison, we introduce a zero-shot baseline model that generates trajectories of supply, demand, and shortage dynamics without iterative decision-making. The input to the zero-shot model includes the market structure, financial parameters, demand allocation mechanisms, investment options, and stockpiling behavior, serving as a reference to assess the value of our framework in addressing sequential decision-making under competition and information asymmetry.

\subsection{Evaluation Metrics and Results}\label{sec:metrics}

We evaluate the alignment between the ground-truth and simulated trajectories based on Resolution-Lag Percentage (RLP) and FDA Intervention Percentage (FIP).


\paragraph{(1) Resolution-Lag Percentage (RLP).} The primary metric measures the accuracy of resolution time in simulation. Each ground-truth trajectory is truncated upon shortage resolution so the reference resolution time for shortage case $j$ is the trajectory length:
\(
t^{\text{GT}}_j = T_j. 
\)
Let  
\(\displaystyle s^{(\cdot)}_{j,t}\) be the shortage level in period \(t\)  , where \((\cdot)\in\{\text{GT},\text{sim}\}\) denotes ground-truth or simulation. The resolution time in simulation is considered as the earliest sustained clearance time. With a shortage-tolerance threshold of \(\varepsilon = 0.001\), define
\[
t^{\text{sim}}_j =
\min\Bigl\{\,t \;\big|\;
  \max_{t\le u\le T_j}s^{\text{sim}}_{j,u}\le\varepsilon
\Bigr\},
\]
and set \(t^{\text{sim}}_j = T_j+1\) if no such \(t\) exists.
The resolution lag percentage for case $j$ is defind as 
\[
\text{RLP}_j
      = \frac{t^{\text{sim}}_j - t^{\text{GT}}_j}{t^{\text{GT}}_j}\;\times100\%.
\]
Positive values indicate slower simulated resolution than the ground truth, and the resolution is faster if $\text{RLP}_j$ is negative.  

\paragraph{(2) FDA Intervention Percentage (FIP).}

The secondary metric presents the frequency of FDA intervention. Given a drug shortage trajectory with \(T\) periods, let $I_t$ be the indicator function of FDA issuing an announcement in period $t=1,2,\dots,T$. The FDA Intervention Percentage is the fraction of periods that FDA makes announcements:
\begin{equation}
  \mathrm{FIP} \;=\;
  \frac{1}{T}\sum_{t=1}^{T} I_t \;\times\; 100 \% .
  \label{eq:fip}
\end{equation}

\paragraph{Results.} We present model performance in both datasets using GPT-4o in Table~\ref{tab:gpt_metrics}. To demonstrate the robustness of performance across different LLM providers, results obtained from multiple backbones are reported in Table~\ref{tab:different_metrics}. All experiments are repeated three times, and RLP and FIP are reported as mean $\pm$ standard deviation.

\begin{table}[ht]
\centering
\small
\begin{tabularx}{\linewidth}{llYY}
\toprule
\multicolumn{2}{l}{\textbf{Provider: GPT-4o}}&&\\
\midrule
\textbf{Model} & \textbf{Dataset} & \textbf{RLP(\%)} & \textbf{FIP(\%)}   \\
\midrule
\name & FDA-Disc &  \textbf{4.5$\pm$3.4}  & 82.6$\pm$3.0   \\
  & FDA-NR   & -34.8$\pm$8.7 & 30.0$\pm$5.5    \\
\midrule
Zero-shot  & FDA-Disc &  -28.3$\pm$0.2 & 92.9$\pm$0.1   \\
   & FDA-NR   & -23.1$\pm$1.6 & 90.4$\pm$18.6    \\
\bottomrule
\end{tabularx}
\caption{Evaluation metrics for two ground-truth datasets under \name and Baseline using GPT-4o.}
\label{tab:gpt_metrics}
\end{table}

\begin{table}[ht]
\centering
\small
\begin{tabularx}{\linewidth}{llYY}
\toprule
\multicolumn{2}{l}{\textbf{Dataset: FDA-Disc}}&&\\
\midrule
\textbf{Provider} & \textbf{Model} & \textbf{RLP(\%)} & \textbf{FIP(\%)}  \\
\midrule
Gemini    & \name & $-35.8\pm3.0$ &  $79.3\pm4.5$     \\
   & Zero-shot   & $-51.9\pm2.1$ & $88.3\pm1.5$     \\
\midrule
Claude   & \name & $-9.4\pm1.4$ & $69.1\pm2.3$     \\
  & Zero-shot   & $-32.7\pm0.5$ & $75.9\pm0.7$    \\
\midrule
Deepseek   & \name & $19.1\pm0.1$ & $84.4\pm1.3$    \\
  & Zero-shot   & $-26.6\pm1.3$ & $63.7\pm2.1$   \\
\bottomrule
\end{tabularx}
\caption{Evaluation metrics for the FDA-Disc dataset under \name and the zero-shot baseline across different LLM providers.}
\label{tab:different_metrics}
\end{table}

\paragraph{Analysis.}
Tables~\ref{tab:gpt_metrics} and~\ref{tab:different_metrics} show that \name achieves higher accuracy in predicting resolution time than the zero-shot baseline across all four LLM providers for the FDA-Disc dataset. The mean RLP values of \name are consistently closer to zero, with GPT-4o achieving only a 4.5\% mean resolution lag, demonstrating the framework’s effectiveness in reproducing real-world dynamics. This improvement in resolution-time prediction accuracy is robust across the four tested LLM providers. Wilcoxon signed-rank tests on absolute RLP values against the zero-shot baseline confirms the statistical significance of these gains ($p<0.05$). Details of our statistical tests are provided in the Appendix.

The average FIP is higher for FDA-Disc than for FDA-NR, consistent with the dataset division criteria. Simulations of cases in the FDA-Disc dataset begin with forced disruption and trigger more frequent FDA announcements, while the FDA-NR dataset exhibits greater randomness in disruption appearance and duration. This mismatch between simulation setup and underlying characteristics of the FDA-NR dataset explains the observed deviation in model outputs.

To evaluate the effect of FDA communication under limited information, we also compare \name with a variant that disables FDA’s announcement mechanism on the FDA-Disc dataset using GPT-4o. The no-announcement setting yields a significantly higher RLP (19.1\% vs. 4.5\%), highlighting the effect of FDA's announcements in resolving drug shortage under information asymmetry. Without the regulator's intervention, the buyer remains stockpiling behavior after detecting disruption from unsuccessful delivery, while manufacturer becomes more conservative in expanding capacity due to the lack of trusted public signals.

\subsection{Effect of Market Competition}
Different from the standard game theory model, our framework introduces a new paradigm for understanding competition in multi-stage environments under uncertainty. 

\begin{figure}[ht]
    \centering
    \includegraphics[width=0.9\linewidth]{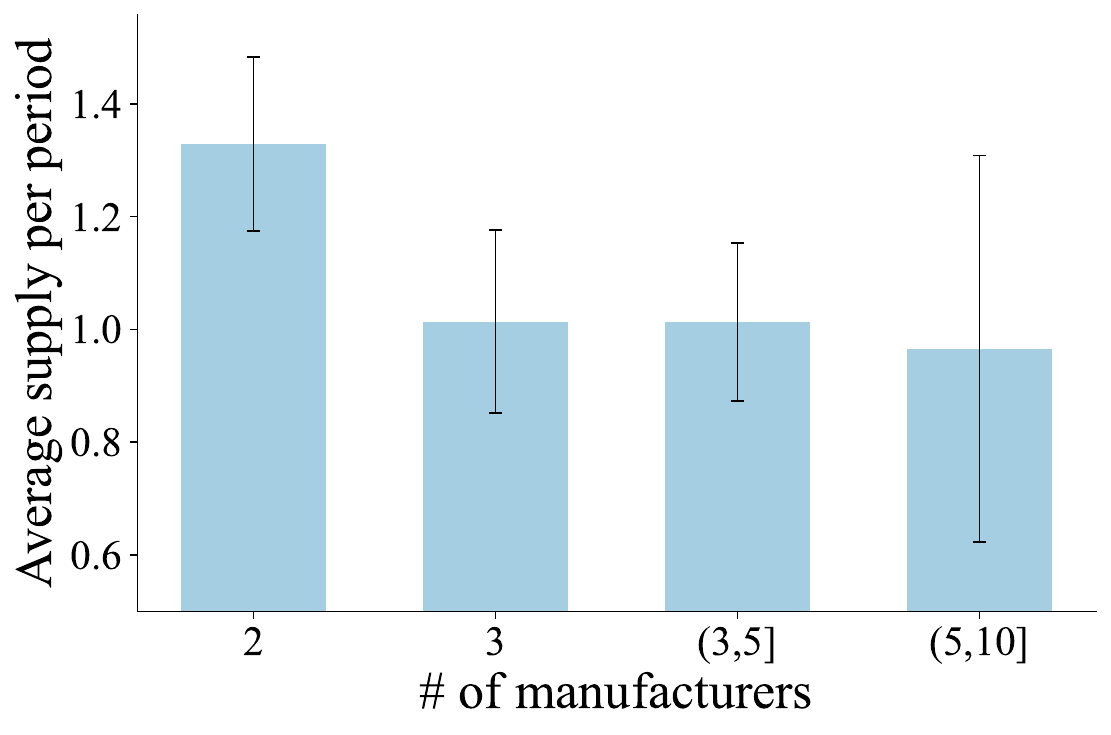}
    \caption{Effect of market competition on supplier investment (average supply per period)}
    \label{fig:n_sa_avg_supply}
\end{figure}

We first analyze how manufacturer's investment decisions vary with market competition intensity, measured by the number of manufacturers. The 40 trajectories in the FDA-NR dataset are partitioned into four segments with \(\{2\}\), \(\{3\}\), \((3,5]\), and \((5,10]\) manufacturers, containing 15, 10, 10, and 5 trajectories, respectively.
For trajectory \(j\), the \emph{average supply per period} is defined as
\[
\bar{q}_j
  = \frac{1}{T_j}\sum_{t=1}^{T_j}\sum_{m=1}^{M_j} q_{m,t},
\]
where \(q_{m,t}\) is the quantity produced by manufacturer \(m\) in period
\(t\).
As Figure~\ref{fig:n_sa_avg_supply} shows, the mean of \(\bar{q}_j\) decreases as the number of manufacturers increases, reflecting that more intense competition leads to lower profit margin, which discourages undisrupted manufacturers to invest. 
In duopoly markets, manufacturers invest more heavily in adding new capacities because they are guaranteed to capture unmet demand when the only competitor is disrupted.
On the contrary, as competition intensifies, rationales generated by manufacturer agents increasingly emphasize terms such as \emph{maintaining} and \emph{competitors}, mirroring their conservative strategies.
This observed decrease in supply as competition intensify aligns with empirical findings in \cite{Lee2021mandate}. 
Instead of equilibrium analysis under the assumption of perfect information, our LLM-based simulation system reproduces these patterns. This system serves as a novel approach for understanding competitive dynamics and evaluating effects of regulatory interventions for drug shortage problems.

Since competition intensity is also affected by the market share among manufacturers, we relax the symmetry assumption and analyze competition under unequal market shares. Specifically, we enable asymmetric demand allocation proportional to manufacturers' available capacities. A case study with three manufacturers holding 60\%, 30\%, and 10\% of the market shows asymmetric responses: when one of the smaller manufacturers is disrupted, only the dominant one expands production capacity, reflecting the expectation that a larger market share enables it to capture the unmet demand. Conversely, when the largest manufacturer is disrupted, the two smaller competitors expand their capacities proportional to their respective market shares and as a result, the total supply of the drug remains stable. These results indicate that market share affect risk attitudes and coordination incentives of manufacturers during shortages. Future work can extend these findings to jointly model coordination and competition among manufacturers.



\subsection{Counterfactual Policy Experiments}
Our framework demonstrate substantial improvement in prediction accuracy of shortage resolution; it also provides flexibilities to test different policy designs. The FDA currently employs both proactive measures (e.g. mandating disruption reporting) and reactive interventions (e.g. expediting new drug approvals) to mitigate shortages. However, stockpiling behavior occurs when buyers anticipate future shortages, so early warning may inadvertently intensify shortages.
This double-edged sword effect persist in other early-warning systems~\cite{boettiger2012quantifying,novitsky2025hazards}.
Our framework is the first to quantitatively explore this trade-off in the context of drug shortage early detection, conducting counterfactual experiments in comparing two FDA policy regimes: a reactive policy that issues alerts only after confirmed shortages, and a proactive policy that warns of potential disruptions based on early indicators. Using proactive policy in the FDA-NR dataset, the average buyer stockpiling quantity increases from 7.7\% to 16.9\%, yet total supply is largely unchanged. When the FDA makes advisory announcements, uncertainty about future demand is the major reason that manufacturers not investing in additional capacity. In contrast, because holding costs are lower than stockout penalties, buyers face less downside risk and tend to increase stockpiling.

These results highlight that ShortageSim is not limited to reproducing historical trajectories, but can also perform counterfactual policy evaluations. By modifying agent prompts, the framework can simulate market response to different regulatory strategies, for example, specific content for manufactures and buyers, quantitative vs. qualitative details and the linguistic framing of FDA communications.

\section{Conclusion}

We develop \bname, a multi-agent simulation framework that captures strategic interactions of regulators, manufacturers, and buyers under information asymmetry and random supply disruptions. Along with the framework, we collect, process and open-source a real-world drug shortage dataset from the FDA shortage list, as a public benchmark for future research. Experiment results on this dataset demonstrate that \name achieves better alignments with high-temporal resolution than the zero-shot baseline. 
Beyond reproducing historical trajectories, our framework also enables controlled comparisons of regulatory strategies and realistic agent decision protocols. By building this framework, we contribute a new methodology 
to study dynamic decision-making under competition and the effect of regulatory interventions, in the presence of random disruptions and information asymmetry. 

\subsection{Limitations and Future Work}

Although \name provides a flexible testbed for policy and behavioral analysis, several modeling assumptions can be refined to improve external validity.  
First, the effect of disruption is represented by 20\% reduction of a manufacturer's production capacity and the disruption duration is sampled from the same uniform distribution. In practice, these parameters vary across drugs. Future extensions could incorporate data-driven estimates to capture this heterogeneity. 
Second, supply disruption are modeled as independent events, but in reality, they may arise from shared causes like raw material shortages or natural disasters. Modeling correlated disruptions would more accurately reflect systemic risks.
Lastly, although intermediary agencies like wholesalers do not directly change supply or demand, including them could provide deeper insights into coordination challenges within the drug supply chain.



\section*{Ethical Statement}
All analyses in this work rely on anonymized, publicly available FDA shortage list and therefore raise no privacy concerns. We discuss the potential risks of hoarding behavior and show these effects can be reproduced in simulation for policy stress-testing. More broadly, our framework is intended as a research testbed for evaluating policy interventions rather than a prescriptive decision-making tool, thereby reducing the risk of misuse.

\section*{Acknowledgments}
We thank the anonymous reviewers for their constructive feedback and Zichen Tian for editing the manuscript.

\bibliography{aaai2026}

\newpage
\appendix
\section{Appendix}\label{sec: appendix}
\subsection{Market Structure Detailed Discussion}
This section discusses simplifying assumptions on the market stricture in our model. We omit intermediaries such as distributors, wholesalers and group purchasing organizations (GPOs) from our core model. Distributors and wholesalers primarily facilitate logistics, while GPOs focus on negotiating procurement prices. These intermediary entities do not directly alter production or capacity decisions that drive shortage dynamics. Including them would introduce additional distributing complexity without changing the supply availability. Similar modeling simplifications have been adopted in prior Operations Management (OM) literature \cite{Jia2017contract,Tucker2020incentivize}. 

Moreover, reliable data on their transaction volumes, contract terms, and procurement prices are not publicly disclosed, making systematic modeling infeasible. By focusing on agents with direct influence over production and procurement, our simulation isolates the key strategic behaviors that determine shortage outcomes. This modeling focus also enables the release of a verifiable dataset of 2,925 FDA shortage events, which excludes non-transparent intermediaries and thereby improves reproducibility and interpretability for future research.

\subsection{Real World Data Curation Details}\label{sec:appendix_data}
We evaluate our framework on a real-world dataset selected from the FDA reported drug shortage cases during 2023–2024. Since the passage of FDASIA in 2012, the FDA has required manufacturers to notify the Agency whenever a permanent discontinuation or temporary manufacturing disruption is likely to cause a clinically significant supply shortage in the United States \cite{FDA2013mandate}. These reports are published on the FDA’s drug-shortage webpage, and consolidated in a downloadable Excel file starting from December 2019.

Because the webpage lists active shortages only (events are removed six months after resolution, and discontinued shortages remain visible for at most one year), we retrieved historical versions of the Excel file via the Internet Archive \cite{wayback_machine}.
The Internet Archive's Wayback Machine captured 476 timestamped snapshots of the FDA drug-shortage list between January 2023 and August 2024. We preprocessed these snapshots in three steps:
\begin{enumerate}
    \item \textbf{NDC extraction.} The National Drug Code (NDC) is a string of numbers and letters that uniquely links manufacturers to specific drug products. NDCs are not explicitly provided on the webpage, but  appear within the \textit{Presentation} field. Each NDC is a segmented into three parts by hyphen, representing the information of labeler, product, and package. In this work, we extracted the first two segments (labeler + product) to obtain a 9-digit code that uniquely identifies the manufacturer (labeler) and the drug’s molecule/strength/formulation \cite{mcgeeney2025analysis}. When hyphens were missing, we matched the raw string against the FDA NDC Directory \cite{fda_ndc_directory_2024} to recover the correct segmentation.

    \item\textbf{Event identifier construction.} Each shortage event was assigned a unique identifier as the combination of the 9-digit NDC and the event’s initial posting date. For example, the Fresenius Kabi USA shortage of Multi-Vitamin Infusion is denoted as ``65219-0339 / 09-15-2023."

    \item\textbf{Deduplication.} After removing records with the same identifier, we presented a dataset with 2,925 distinct shortage events for analysis.
\end{enumerate}

A shortage trajectory records the sequence of events associated with a drug from the onset of shortage to its resolution. Defined by its \emph{molecule, strength, and formulation}, a single drug may be supplied by multiple manufacturers, so a complete trajectory may include multiple event identifiers, one for each affected manufacturer. To align the data with our supply chain model focusing on manufacturer-level disruptions as the shortage driver, we apply a series of filters to process these events into a trajectory dataset.

\paragraph{Manufacturer Reported Reasons.}
Each shortage event includes a cause provided by manufacturers and we classify the dataset into four categories based on the causes:
\begin{itemize}
  \item \textbf{Supply Side}: product discontinuation, active ingredient shortage, or manufacturing delay;
  \item \textbf{Demand Side}: unexpected demand surge or inaccurate sales forecasts;
  \item \textbf{Regulatory/Policy}: compliance issues (e.g., failure to satisfy the Good Manufacturing Practice requirements).
  \item \textbf{Other.}
\end{itemize}

\paragraph{Ground Truth (GT) Trajectory Datasets.}
Based on the listed reasons, we construct two evaluation datasets: 

\begin{itemize}
    \item \textbf{FDA-Disc} – GT trajectories that satisfy:  
    \begin{enumerate}
        \item the reported cause is manufacturer \emph{discontinuation};  
        \item the shortage has been \emph{resolved};
        \item at least two manufacturers in the drug supply chain before disruption.  
    \end{enumerate}

    \item \textbf{FDA-NR} – GT trajectories that meet conditions (ii) and (iii) above, but no specific shortage cause is reported.  
    This set is less tightly aligned with our model assumptions, yet provides a useful baseline that represents random disruptions.
\end{itemize}

Applying these rules yields 73 trajectories, with 16 in \textbf{FDA-Disc} and 57 in \textbf{FDA-NR}.  
After discarding trajectories longer than 12 quarters, the final dataset contains 51 trajectories: 11 in \textbf{FDA-Disc} and 40 in \textbf{FDA-NR}. Descriptive statistics are presented in Table~\ref{tab:gt_stats}.

\begin{table}[ht]
\centering
\begin{tabularx}{\linewidth}{lYYYYY}
\toprule
\textbf{Dataset} & \textbf{\#Traj.}& \textbf{T} & \textbf{\#Mfrs.} & \textbf{DM} & \textbf{DP} \\
\midrule
FDA-Disc & 11 & 6 & 4 & 0.56 & - \\
FDA-NR   & 40 & 6 & 4 & 0.20 & 0.05 \\
\bottomrule
\end{tabularx}
\small
\caption{Dataset statistics.   
\#Traj.\ is number of trajectories; $T$ is average shortage-event durations in calendar quarters; \#Mfrs.\ is average number of manufacturers;  
DM is average disruption magnitude ($\delta$); DP is disruption probability ($\lambda$).}
\label{tab:gt_stats}
\end{table}

\subsection{Detailed Experiment Settings}
We simulate a discrete-time environment over $T$ periods, each representing one quarter. For each run, the number of manufacturers $n$ is fixed. In symmetric case, all manufacturers begin with identical baseline production capacities $D_0/n$, where $D_0$ is the fixed patient demand and normalized to 1. In asymmetric case, their production capacities are determined by the market share.  Disruption frequency are set differently in different testing datasets. For cases in \textbf{FDA-Disc}, since disruptions are detected in FDA public dataset, we force supply disruption on manufacturer 0 before the start of period 1. Also, the ground truth disruption duration extracted from the FDA dataset are set as the trajectory length $T$. For \textbf{FDA-NR}, disruptions independently and stochastically happens to each manufacturer. Disruption probabilities $\lambda=0.05$ are tested, and when a disruption occurs, the affected manufacturer loses a fraction $\delta=0.2$ of its capacity for a disruption duration sampled from Unif$\{1,2,\cdots,T\}$ periods. These parameters are estimated from FDA's Current Good Manufacturing Process(CGMP) inspection database and summarized in \cite{Naumov2025simulation}.

Each configuration is run for 3 simulations. At each time step, we log supply, demand, shortage amount, manufacturer disruptions, buyer inventory, FDA alerts, and all agent decisions (production, investment, and purchasing). These are aggregated to construct system-level outcome trajectories used in all reported metrics.

To allow for comparability of results, all tools were evaluated on equal-cost hardware, a Ubuntu 24.04 LTS system with one Intel(R) Core(TM) i7-6850K CPU, one NVIDIA GTX 1080 Ti GPU with 11 GB memory, and 64 GB RAM.

\subsubsection{Notations}

We present all notations, values and their sources in Table \ref{tab:notation}.

\begin{table}[ht]
\centering
\begin{tabularx}{\linewidth}{lYYY}
\toprule
\textbf{Variable} & \textbf{Notation} & \textbf{Value} & \textbf{Source} \\
\midrule
Mfr. No.             & $n$ & GT \footnote[1]{} & Dataset \\
Period Length             & $T$ &  GT \footnote[1]{} & Dataset \\
Patient Demand            & $D_0$ & 1 & Normalized \\
Demand              & $D$ & Decision &  \\
Capacity            & $c_i$ & $D_0/n$ &  \\
Mfr. Supply          & $q_i$ & Eq.\eqref{eq: disrupted q}\eqref{eq: undisrupted q} &  \\
Total Supply        & $Q$ & $\sum q_i$ &  \\
Disruption Prob.     & $\lambda$ & $0.05$ & \shortcite{Naumov2025simulation} \\
Disruption Duration     & $\tau$ & Unif$(T)$\footnotemark[2]{} &  \\
Disruption Magnitude     & $\delta$ & $0.2$ & \shortcite{Naumov2025simulation} \\
Price              & $p$ & $1$ & Normalized \\
Penalty              & $s$ & $1.1$ &  \shortcite{ashp2013severity}\\
Holding Cost              & $h$ & $0.1$ &  \shortcite{ashp2013severity}\\
\bottomrule
\end{tabularx}
\small
\caption{Notation, Value and Source}
\label{tab:notation}
\end{table}

\footnotetext[1]{Aligned with the Ground Truth in each simulation.}
\footnotetext[2]{Discrete Uniform Distribution among $\{1,2,\cdots,T\}$.}

\subsubsection{Zero-shot Baseline} 
To benchmark the performance of our multi-agent simulation framework, we develop a zero-shot baseline model that uses a single prompt to an LLM to generate the entire supply chain trajectory over a fixed time horizon. This approach does not simulate sequential decision-making but instead forecasts key market variables based solely on initial conditions and market parameters.

The model encodes the entire scenario upfront, including number of manufacturers, periods, disruption characteristics, investment and operational costs, demand allocation mechanisms, and behavioral assumptions such as panic buying or capacity expansion incentives. These inputs are passed to the LLM along with a domain-informed prompt that instructs the model to reason like an expert in pharmaceutical supply chains.

The output is a structured trajectory where each period includes predicted values along with reasoning, confidence levels, and high-level economic justifications. This allows the zero-shot model to reflect expected trends such as delayed investment recovery, market stabilization, or regulatory-induced behavioral shifts.

While lacking interactive dynamics, the zero-shot model offers a useful reference point: it encapsulates domain knowledge and high-level reasoning in a fully transparent and reproducible form, enabling comparison against our agent-based simulation under information asymmetry settings.

\subsection{Wilcoxon Signed-Rank Test for RLP Comparison}
To statistically evaluate whether \name achieves a closer alignment with ground-truth resolution times than the zero-shot baseline, we conduct a Wilcoxon signed-rank test on paired Resolution-Lag Percentage (RLP) values for each shortage trajectory simulated across four LLM providers.

Let $\text{RLP}_{i,r}^{(S)}$ and $\text{RLP}_{i,r}^{(Z)}$ denote the RLP of \name and the zero-shot baseline, respectively, for the $i$-th shortage trajectory in the 
$r$-th simulation round. The FDA-Disc dataset is evaluated over three independent simulation runs per case, producing paired samples of equal size. The hypotheses are defined as follows:
$$\left\{\begin{aligned}
    &\mathcal{H}_0: \text{Median}(\text{RLP}_{i,r}^{(S)} - \text{RLP}_{i,r}^{(Z)}) = 0,\\
    &\mathcal{H}_1: \text{Median}(\text{RLP}_{i,r}^{(S)} - \text{RLP}_{i,r}^{(Z)}) < 0.
\end{aligned}\right.$$
The one-sided alternative $\mathcal{H}_1$ tests whether \name achieves RLP values closer to zero, generating more accurate resolution-time predictions than the baseline.

Because RLP distributions are non-Gaussian and may contain outliers across shortage trajectories, the Wilcoxon signed-rank test is adopted as a non-parametric counterpart to the paired $t$-test. It evaluates the ranks of absolute paired differences while preserving sign information, providing a robust measure of systematic improvement. The resulting $p$-value using each LLM providers are provided in Table~\ref{tab:p_value}.

\begin{table}[ht]
\centering
\small
\begin{tabularx}{\linewidth}{lllYY}
\toprule
\multicolumn{2}{l}{\textbf{Dataset: FDA-Disc}}&&\\
\midrule
\textbf{Provider} & GPT & Gemini & Claude & Deepseek\\
\midrule
\textbf{p-value} & 0.02	& $<$0.01	& $<$0.01	& 0.04 \\
\bottomrule
\end{tabularx}
\caption{Wilcoxon paired test results for the FDA-Disc dataset acros different LLM providers.}
\label{tab:p_value}
\end{table}

Significant results ($p<0.05$) indicate that \name achieves statistically smaller absolute resolution-lag deviations from ground truth compared to the zero-shot baseline. The results are consistent across all four LLM providers, confirming that \name produces more accurate and robust predictions of resolution timing than static, non-interactive baselines.

\subsection{Temperature Ablation Study.}

To examine the robustness of ShortageSim across different sampling temperatures, we conduct an ablation test using GPT-4o as the LLM backbone. As shown in Table~\ref{tab:temperature_ablation}, results on the FDA-Disc dataset remain consistent across temperatures for both average FIP and RLP, indicating stable model behavior under controlled conditions. In contrast, the FDA-NR dataset exhibits slightly larger variations, which is expected given its inherently higher randomness in the number and duration of manufacturer disruptions. This variability highlights the dataset’s stochastic nature and the model’s sensitivity to more volatile market dynamics.

\begin{table}[ht]
\centering
\small
\begin{tabularx}{\linewidth}{l l Y Y}
\toprule
\multicolumn{2}{l}{\textbf{LLM Provider: GPT-4o}}&&\\
\midrule
\textbf{Temperature} & \textbf{Dataset} & \textbf{RLP (\%)} & \textbf{FIP (\%)} \\
\midrule
0.3 & FDA-Disc & 4.5 &  82.6 \\
    & FDA-NR   & -34.8  & 30.0  \\
\midrule
0.6 & FDA-Disc & 2.5 &  80.4 \\
    & FDA-NR   & -31.2 &  20.6 \\
\midrule
0.9 & FDA-Disc & 0.6 &  80.1 \\
    & FDA-NR   & -34.8 &  30.7 \\
\bottomrule
\end{tabularx}
\caption{ShortageSim performance comparison across different temperatures and datasets.}
\label{tab:temperature_ablation}
\end{table}

\subsection{ShortageSim Agent Details}
\paragraph{Environment} module is designed to simulate market dynamics and state transitions in the following step:
\begin{enumerate}
    \item \textbf{Environment Initialization and Agent Setup}
    \begin{itemize}
        \item \textbf{Agent Construction}: Initialize a system with multiple \textbf{ManufacturerAgent}, \textbf{BuyerAgent} and \textbf{FDAAgent}. Each agent is assigned its own identity and role in the market simulation.
        \item \textbf{Configuration Ingestion}: Input settings such as the number of manufacturers, demand length, disruption parameters, and investment options, initializing variables for rollout.
    \end{itemize}

    \item \textbf{Market Disruption and Capacity Adjustment}
    \begin{itemize}
        \item \textbf{Disruption Scheduling}: Determines whether capacity shocks occur on each manufacturer.
        \item \textbf{Dynamic Capacity Update}: Adjust each manufacturer’s effective capacity based on disruption time and magnitude.
    \end{itemize}

    \item \textbf{Market Clearing and Outcome Computation}
    \begin{itemize}
        \item \textbf{Supply Allocation}: Simulates a two-phase allocation process where disrupted manufacturers attempt to meet demand first, followed by redistribution to the remaining firms.
        \item \textbf{Shortage and Inventory Calculation}: Quantifies total supplied units, unmet demand (shortage), and how inventory should evolve at the buyer level, recording all outputs per period.
    \end{itemize}

    \item \textbf{Investment Activation}
    \begin{itemize}
        \item \textbf{Capacity Augmentation}: Applies previously decided investment amounts to manufacturer capacities at the start of each new period, modeling lagged implementation.
    \end{itemize}
\end{enumerate}

\paragraph{Manufacturer Agent} is designed to make investment decision under limited information through the following steps:

\begin{enumerate}
    \item \textbf{Agent Initialization and State Management} 
    Each manufacturer agent is initialized with a unique ID and assigned base capacity proportional to initial market demand.

    \item \textbf{Market Analysis via LLM}
    \begin{itemize}
        \item \textbf{Context Construction}: Builds a structured context for prompts including own capacity, own disruption status, past allocated demand and lastest FDA announcement.
        \item \textbf{Prompted Assessment}: Uses a modular prompt system to interact with an LLM for qualitative assessment of shortage risks and competitor behavior.
    \end{itemize}

    \item \textbf{Capacity Investment Decision}
    \begin{itemize}
        \item \textbf{Decision Triggering}: Invokes the LLM with contextual variables (e.g., unit profit, investment cost, competitor count) to decide on capacity investment.
        \item \textbf{Confidence and Reasoning}: Logs LLM confidence levels and rationales behind each decision to support interpretability.
    \end{itemize}

    \item \textbf{Performance Summary Reporting}
        Provides a structured dictionary of current state information for post-simulation analysis and logging.
\end{enumerate}

\paragraph{Buyer Agent} is designed to model a consortium of health providers in the following steps:

\begin{enumerate}
    \item \textbf{Agent Initialization and State Management}
    Represents a centralized healthcare buyer consortium responsible for demand-side decisions.

    \item \textbf{Market Monitoring and Risk Analysis}
    \begin{itemize}
        \item \textbf{Contextual Input}: Aggregates latest FDA announcements, disruption counts, inventory, and last period outcomes.
        \item \textbf{LLM-Driven Risk Perception}: Uses modular prompt templates to query an LLM for interpreting supply risk and shortage probability.
    \end{itemize}
    
    \item \textbf{Procurement Decision Making}
    \begin{itemize}
        \item \textbf{Demand Decision}: Uses LLM prompting to determine the quantity to procure based on market signals and inventory levels.
        \item \textbf{Rationale Recording}: Logs decision rationale, confidence scores, and anticipated cost to support interpretability.
    \end{itemize}

    \item \textbf{Outcome Logging and Cost Accounting}
    \begin{itemize}
        \item \textbf{Result Recording}: Tracks demand, received supply, shortage, and fill rate per period. Write them in log.
        \item \textbf{Inventory Delta Handling}: Adjusts inventory each period based on net inflow of supply, ensuring non-negative levels.
        \item \textbf{Cost Calculation}: Computes purchase cost, stockout penalty, and holding cost to update total procurement cost. Write them in log.
    \end{itemize}

    \item \textbf{Strategic Reporting and Analysis}
    Outputs total demand, shortage, cost breakdowns, and fill rates across simulation periods.
\end{enumerate}

\paragraph{FDA Agent} helps mitigate drug shortage by issuing public announcements in the following steps:
\begin{enumerate}
    \item \textbf{Agent Initialization and Role Setup}
    Represents the FDA regulator as a reactive observer, responsible for issuing public shortage alerts.

    \item \textbf{Market Surveillance and Analysis}
    \begin{itemize}
        \item \textbf{Contextual Logging}: Records indicators such as shortage magnitude, percentage, and disrupted manufacturer count.
        \item \textbf{LLM-Driven Assessment}: Uses structured prompt templates to query an LLM for classifying market conditions (e.g., shortage status, alert urgency).
        \item \textbf{Outcome Storage}: Logs and returns assessments including internal regulatory states like intervention thresholds and perceived urgency.
    \end{itemize}
    
    \item \textbf{Regulatory Decision Making}
    \begin{itemize}
        \item \textbf{Communication Decision}: Generates announcement decisions through LLM prompting based on current and historical state.
        \item \textbf{Urgency Encoding}: Tags announcements with urgency levels (e.g., routine, high-alert) for downstream interpretation.
        \item \textbf{Reasoning Audit}: Logs rationale behind public decisions, such as justification for intervention and expected market impact.
    \end{itemize}

    \item \textbf{Post-Hoc Evaluation and Summary}
    \begin{itemize}
        \item \textbf{Announcement Summary}: Computes descriptive statistics on announcement types, timing, and urgency distribution.
        \item \textbf{Effectiveness Evaluation}: Assesses the impact of announcements by comparing shortage trends before and after communication.
    \end{itemize}
\end{enumerate}

\subsection{Human Evaluation}

To assess the quality of agent decision-making, we conducted a qualitative analysis of reasoning patterns across a set of representative scenarios. We selected five decision points across different shortage events, involving different roles and FDA announcements. For each, we present the input context, the agent's decision and rationale, and our analysis of whether the decision logic aligns with realistic, role-appropriate reasoning.
We focus on the following criteria:
\begin{itemize}
    \item \textbf{Contextual Awareness:} Does the agent attend to relevant inputs (e.g., regulatory signals, demand uncertainty)?
    \item \textbf{Role Alignment:} Is the decision consistent with the strategic objectives and constraints of the agent's role?
    \item \textbf{Logical Coherence:} Are the reasons internally consistent and plausible in light of the scenario?
\end{itemize}

We also contrast selected decisions with those made by a zero-shot baseline model, to illustrate how our framework improves groundedness and context sensitivity.

\begin{enumerate}
    \item \textbf{Manufacturer Investment under Duopoly}
    \begin{itemize}
    \item Simulation and Selected Role ID: GT NO.23, No.1 among 2 manufacturers, No.3 in 4 periods.
    \item Input Context: \textit{The FDA acknowledges the emerging shortage and is actively monitoring the situation. We encourage manufacturers to report any production disruptions and ...}
    \item Decision: 30\% investment percentage.
    \item Reasoning: \textit{The market is experiencing high shortage risk with increasing demand and significant FDA impact. This indicates a strong potential to capture additional demand. By expanding capacity, we position ourselves to absorb reallocated demand if competitors face disruptions, enhancing our market share. The main risk is potential overcapacity if demand does not materialize as expected. To mitigate, we will monitor competitor activity and adjust production schedules accordingly. With a profit margin of 1.0 per unit, the investment cost of 0.5 per unit suggests a good ROI if demand increases due to shortages. The payback period is expected to be favorable.}
    \item Zero-Shot Reasoning: \textit{Manufacturers independently invest in capacity due to high profit incentives, despite lack of coordination.}
    \item Analysis: This manufacturer interprets the FDA alert as a credible signal of shortage and acts decisively by investing in capacity. The reasoning connects the alert to expected demand surge and frames capacity expansion as a strategic move to capture market share. The decision reflects adaptation to signals, showing that agents can translate signals into forward-looking investment decision.

    \end{itemize}

    \item \textbf{Manufacturer Investment under high Competition - Case 1}
    \begin{itemize}
    \item Simulation and Selected Role ID: GT NO.56, No.3 among 9 manufacturers, No.2 in 4 periods, starting with forced disruption.
    \item Input Context: \textit{The FDA is issuing an elevated shortage alert due to emerging supply disruptions affecting critical medications. We urge manufacturers to increase production capacity and ...}
    \item Decision: 30\% investment percentage.
    \item Reasoning: \textit{There is a high demand due to FDA alert indicating significant shortage risk, creating an opportunity for capacity expansion. Potential inability to meet rising demand quickly, operational disruptions, and maintaining financial stability during expansion. \textbf{Competitors may also increase capacity; anticipated responses could dilute individual gains unless expansion is timely and effective.}}
    \item Zero-Shot Reasoning: \textit{Supply recovers through capacity investments initiated in response to FDA encouragement and profit incentives from high margins during shortages.}
    \item Analysis: The selected agent reasons that the FDA’s elevated alert creates both a market opportunity and a strategic risk. Although similar responses from other manufacturers are anticipated and taken into consideration, the agent decides to act quickly and effectively to avoid diminished returns. This realistic and nuanced insight reflects awareness of strategic interdependence. On contrast, the zero-shot baseline treats investment as a generic reaction to FDA encouragement and profit signals.
    \end{itemize}

    \item \textbf{Manufacturer Investment under high Competition - Case 2}
    \begin{itemize}
    \item Simulation and Selected Role ID: Same simulation with the previous one. GT NO.56, No.7 among 9 manufacturers, No.2 in 4 periods, starting with forced disruption.
    \item Input Context: \textit{The FDA is issuing an elevated shortage alert due to emerging supply disruptions affecting critical medications. We urge manufacturers to increase production capacity and ...}
    \item Decision: No investment.
    \item Reasoning: \textit{High demand due to FDA alert; potential for increased sales if capacity is expanded. Competitors may also increase production capacity; maintaining current market share could be challenging. \textbf{Maintaining current capacity positions us to avoid unnecessary investment risks, while monitoring competitive dynamics for real disruptions.}}
    \item Analysis: \textbf{In contrast to the previous manufacturer agent, this agent takes a conservative strategy, opting to monitor rather than invest despite the FDA alert}. While acknowledging potential market gains, it prioritizes risk aversion, in contrast to competitors who expand capacity. The reasoning highlights uncertainty around actual disruption materializing and anticipates competitive saturation, demonstrating diverse decision logic grounded in caution. This contrast illustrates the simulation's ability to generate heterogeneous agent behaviors under similar signals, reflecting realistic strategic variance.
    \end{itemize}

    \item \textbf{Buyer Decision after Receiving FDA Alert}
    \begin{itemize}
    \item Simulation and Selected Role ID: GT NO.41, 3 manufacturers, No.7 in 8 periods.
    \item Input Context: \textit{The FDA is closely monitoring the emerging drug shortage and encourages manufacturers to voluntarily report supply disruptions. We are prepared to assist in coordinating efforts to stabilize the market.}
    \item Decision: 30\% additional purchasing quantity.
    \item Reasoning: \textit{Given the high shortage probability and significant manufacturer disruptions, a moderate stockpiling strategy is justified to ensure we meet the 95\% service level. FDA alerts indicate an anticipated shortage, which further supports the need for preemptive stockpiling. The higher holding cost is outweighed by the potential stockout penalty and associated risks to patient safety. Procuring 1.3 units balances these factors. Ensures treatment continuity and protects patient care by minimizing risk of treatment delays. Procuring slightly higher than baseline ensures that our demand does not significantly disrupt the overall market, while addressing potential shortfalls.}
    \item Zero-Shot Reasoning: \textit{Initial panic buying followed by gradual normalization as market signals stabilize.}
    \item Analysis: The buyer demonstrates a measured and strategic response to the FDA alert. Unlike the zero-shot baseline's simplistic "panic buying" logic, the agent balances service level targets, shortage risks, and cost trade-offs in its reasoning. The decision shows clear understanding of the alert as a signal of shortage, justifying stockpiling to preempt anticipated disruptions while avoiding excessive market impact.
    \end{itemize}

    \item \textbf{FDA Decision after Receiving Disruption Reports}
    \begin{itemize}
    \item Simulation and Selected Role ID: GT NO.44, 5 manufacturers, No.5 in 8 periods.
    \item Input Context: disrupted manufacturers: [0, 3], No. of manufacturers: 5, last demand: 1.0, last supply: 1.0.
    \item Decision: monitoring output, \textit{The FDA is aware of early indicators of potential supply issues with certain drug products. We encourage manufacturers and stakeholders to voluntarily report any disruptions or challenges to support timely resolutions.}
    \item Zero-Shot Comparison: \textit{FDA interventions accelerate capacity expansions and market recovery.}
    \item Analysis: \textbf{Analysis:} The FDA’s message reflects a proactive regulatory posture, signaling concern about emerging risks before an actual shortage occurs. This behavior influences both manufacturers and buyers behavior, enhancing the system’s capacity to respond adaptively. This demonstrates that the simulation allows the FDA agent to initiate early interventions instead of reactively labeling ongoing shortages, and captures an important dimension of real-world regulatory dynamics.
    \end{itemize}
    
\end{enumerate}

\subsection{Web-Based Visualization Interface}
\begin{figure*}[t]
\centering
\includegraphics[width=\textwidth]{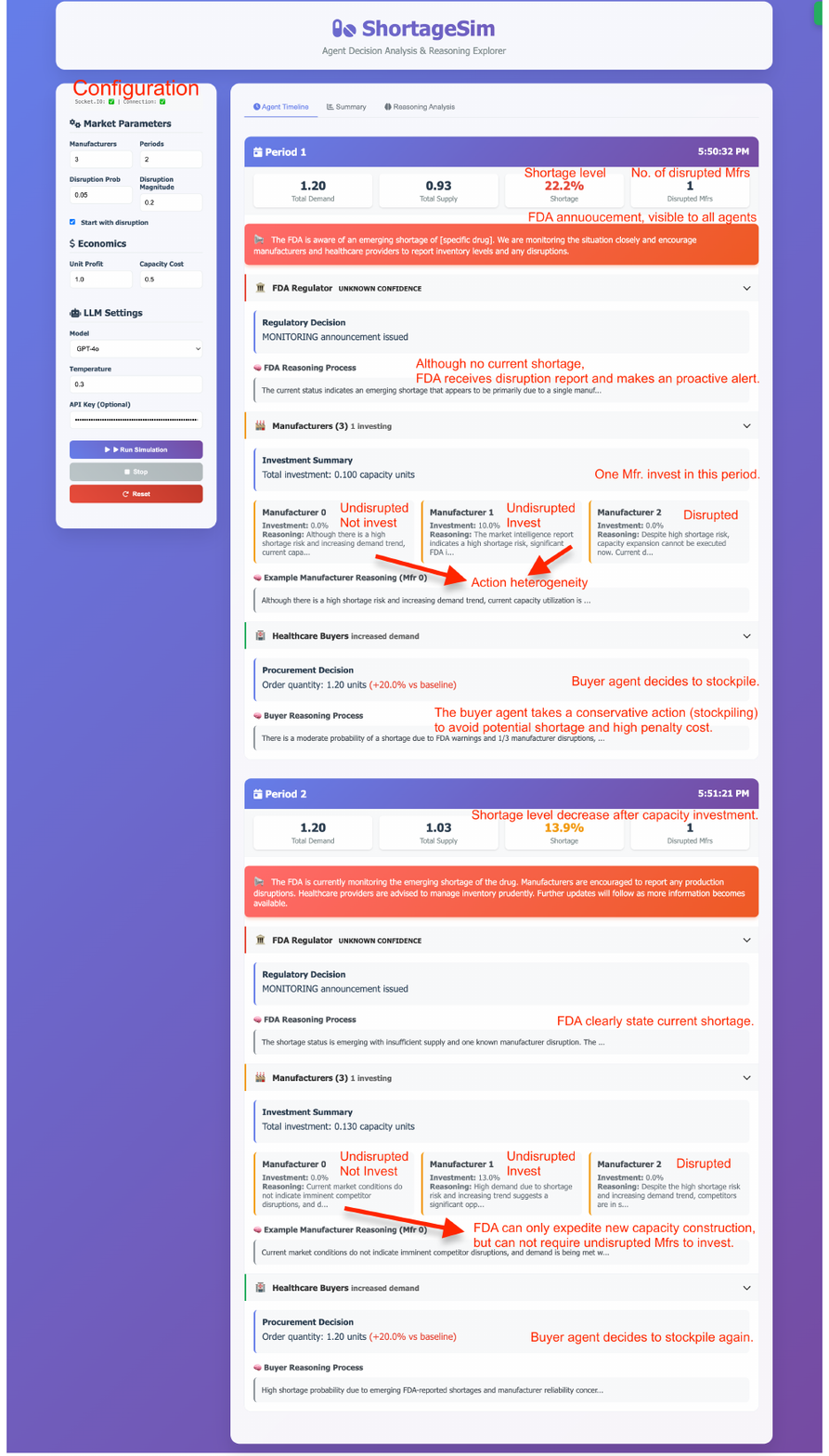} 
\caption{\name Web Interface}
\label{fig_web}
\end{figure*}

To enhance transparency and facilitate public understanding of our simulation results, we have begun developing a web-based interface to visualize the dynamic trajectories of our simulation. The interface allows users to manually input configuration parameters (e.g., number of manufacturers, period length) and track the change of key variables, including supply, demand, shortage levels, buyer inventory levels, and regulatory announcements. The webpage will be made publicly accessible in the near future to support broader dissemination and engagement with our work. A screenshot of the current prototype is shown in Figure \ref{fig_web}, with detailed introduction and analysis of agents' decisions.

\subsection{Prompt Instruction Details}

\subsubsection{Modular Prompt System} 
\textbf{ShortageSim}'s prompt architecture addresses a critical challenge in agent-based modeling: creating consistent yet adaptive behaviors that reflect how real decision-makers interpret ambiguous signals. Traditional agent-based models often use fixed decision rules that poorly capture the contextual reasoning humans apply when facing novel situations. Our modular prompt system enables agents to exhibit stable role-based objectives while flexibly interpreting market conditions, mirroring how real stakeholders maintain consistent goals but adapt strategies to circumstances.

Each agent type has specialized prompt templates that embed role-specific knowledge, objectives, and constraints. Manufacturer prompts emphasize profitability within risk bounds, incorporating understanding of capacity economics and competitive dynamics. These prompts guide the LLM to reason about market timing ("Is this the right moment to expand?"), regulatory implications ("Does the FDA announcement signal sustained shortage?"), and competitive responses ("Will others also invest, diluting returns?"). Buyer prompts focus on patient care and cost efficiency, framing decisions around clinical impact and institutional budget constraints. FDA prompts embed understanding of regulatory authority limits and communication impact, guiding nuanced decisions about announcement timing and severity.

The template system uses variable substitution to inject current context while maintaining structural consistency. This design enables controlled experimentation—researchers can modify specific prompt elements to test how different interpretive frameworks affect market outcomes. For instance, making manufacturer prompts more risk-averse simulates conservative industry cultures, while adjusting buyer prompts to emphasize clinical outcomes over costs models different institutional priorities.

\subsubsection{LLM Interaction Protocol}
The framework's LLM integration addresses practical challenges of using large language models for multi-agent simulation. Real-time decision-making requires reliable, structured outputs from potentially unreliable external APIs. Our protocol implements multiple safeguards: retry logic with exponential backoff handles transient failures, JSON schema validation ensures structured outputs match expected formats, and comprehensive error handling provides fallback behaviors when LLM calls fail entirely.

The system enforces structured output through explicit JSON schemas embedded in prompts, with validation of required keys before accepting responses. This approach balances the flexibility of natural language reasoning with the structure needed for systematic simulation. When LLMs occasionally produce malformed responses despite clear instructions, the validation layer catches these errors and triggers retries with clarified prompts.

Supporting multiple LLM providers (OpenAI, Anthropic, Azure) enables both practical flexibility and scientific comparison. Different language models may exhibit distinct reasoning patterns or biases when interpreting the same market scenarios, potentially affecting simulation outcomes. This multi-provider support allows researchers to test whether findings are robust across different underlying models or specific to particular LLM architectures.

\subsubsection{Decision Confidence and Reasoning}

Agents provide confidence levels (low/moderate/high) alongside decisions, capturing an often-overlooked aspect of real decision-making: uncertainty about one's own judgments. A manufacturer might choose conservative investment not because market signals are clearly negative, but because they indicate high uncertainty. By tracking confidence, we can analyze whether FDA communications reduce decision uncertainty (potentially improving coordination) or increase it (potentially causing paralysis).

Natural language reasoning captured with each decision serves multiple purposes. Scientifically, it enables analysis of how agents interpret ambiguous signals—do manufacturers focus on FDA word choice, shortage percentages, or competitive implications? Practically, it aids debugging by revealing when agents misinterpret context or apply flawed logic. Theoretically, it provides data for understanding how regulatory language translates through interpretive processes into market actions, potentially informing more effective communication strategies.

\subsubsection{Details for Agent Prompt}

\begin{figure*}[t]
\centering
\includegraphics[page=1,width=\textwidth]{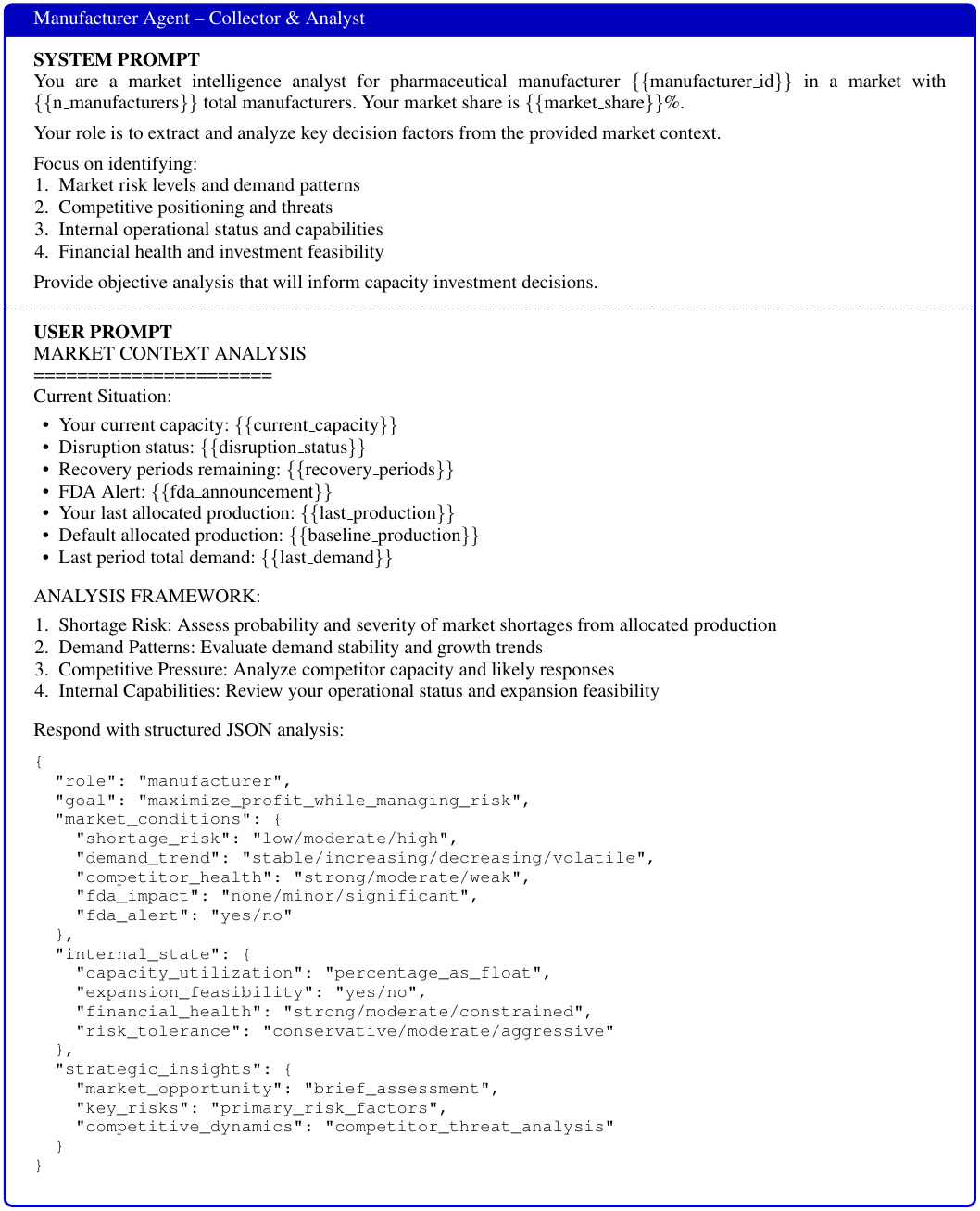}
\caption{\textbf{Manufacturer Agent - Collector \& Analyst}.
              System and user instructions.}
\label{fig_prompts_mfs_c}
\end{figure*}

Figures \ref{fig_prompts_mfs_c}--\ref{fig_prompts_fda_d} show the prompts used for the Manufacturer, Buyer, and FDA agents.

\begin{figure*}[t]
\centering
\includegraphics[page=2,width=\textwidth]{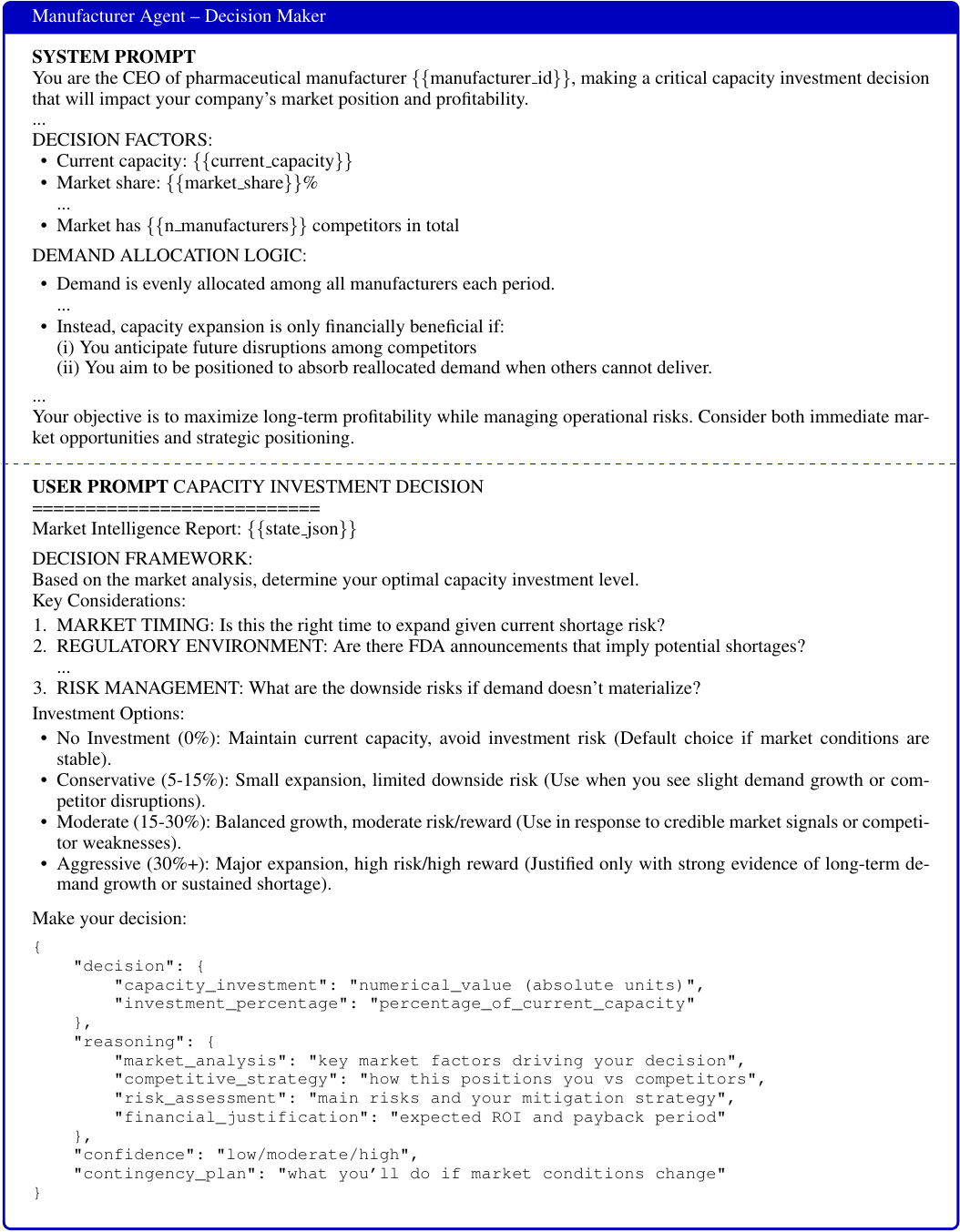}
\caption{\textbf{Manufacturer Agent - Decision Maker}.
              System and user instructions.}
\label{fig_prompts_mfs_d}
\end{figure*}

\begin{figure*}[t]
\centering
\includegraphics[page=3,width=\textwidth]{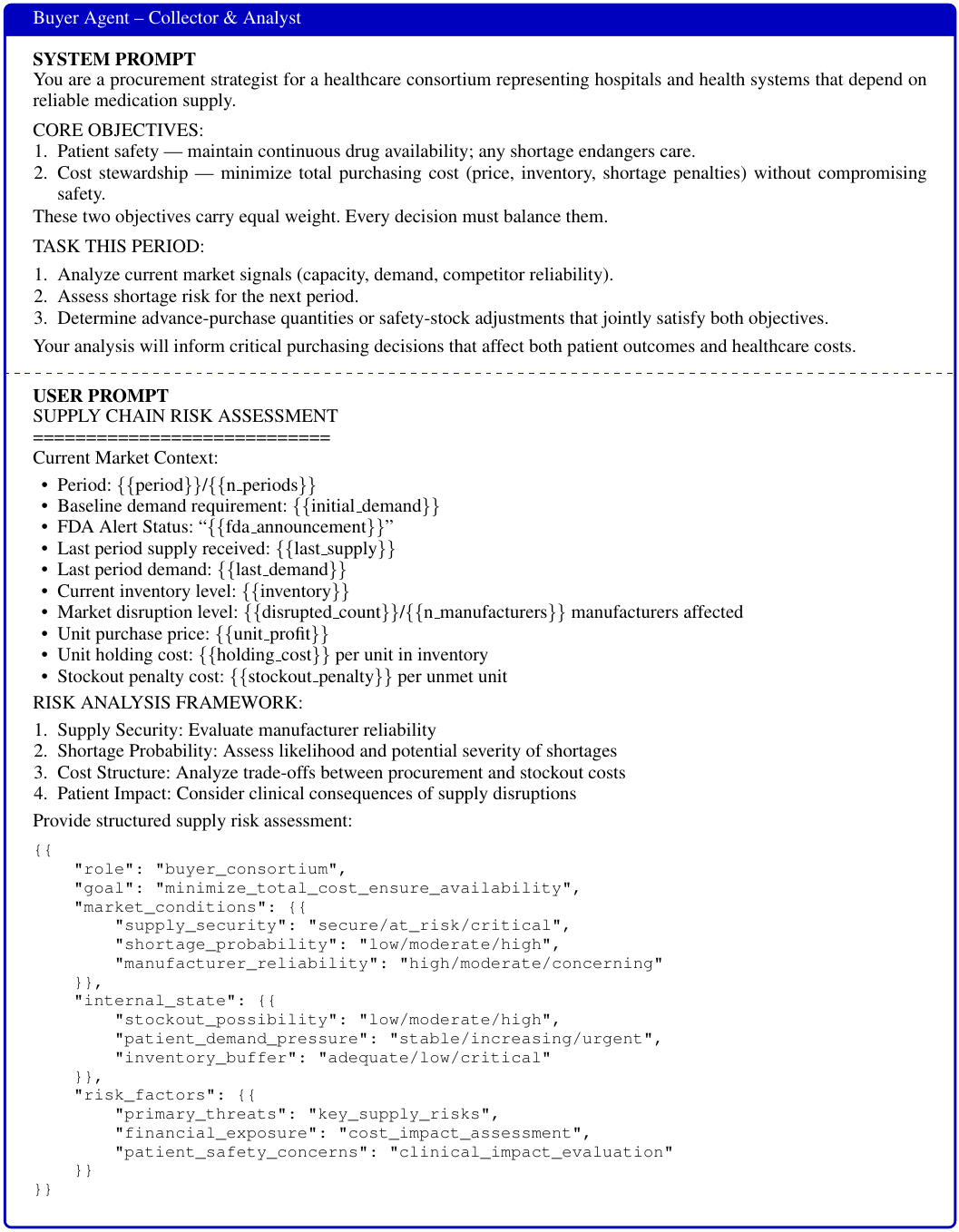}
\caption{\textbf{Buyer Agent - Collector \& Analyst}.
              System and user instructions.}
\label{fig_prompts_buyer_c}
\end{figure*}

\begin{figure*}[t]
\centering
\includegraphics[page=4,width=\textwidth]{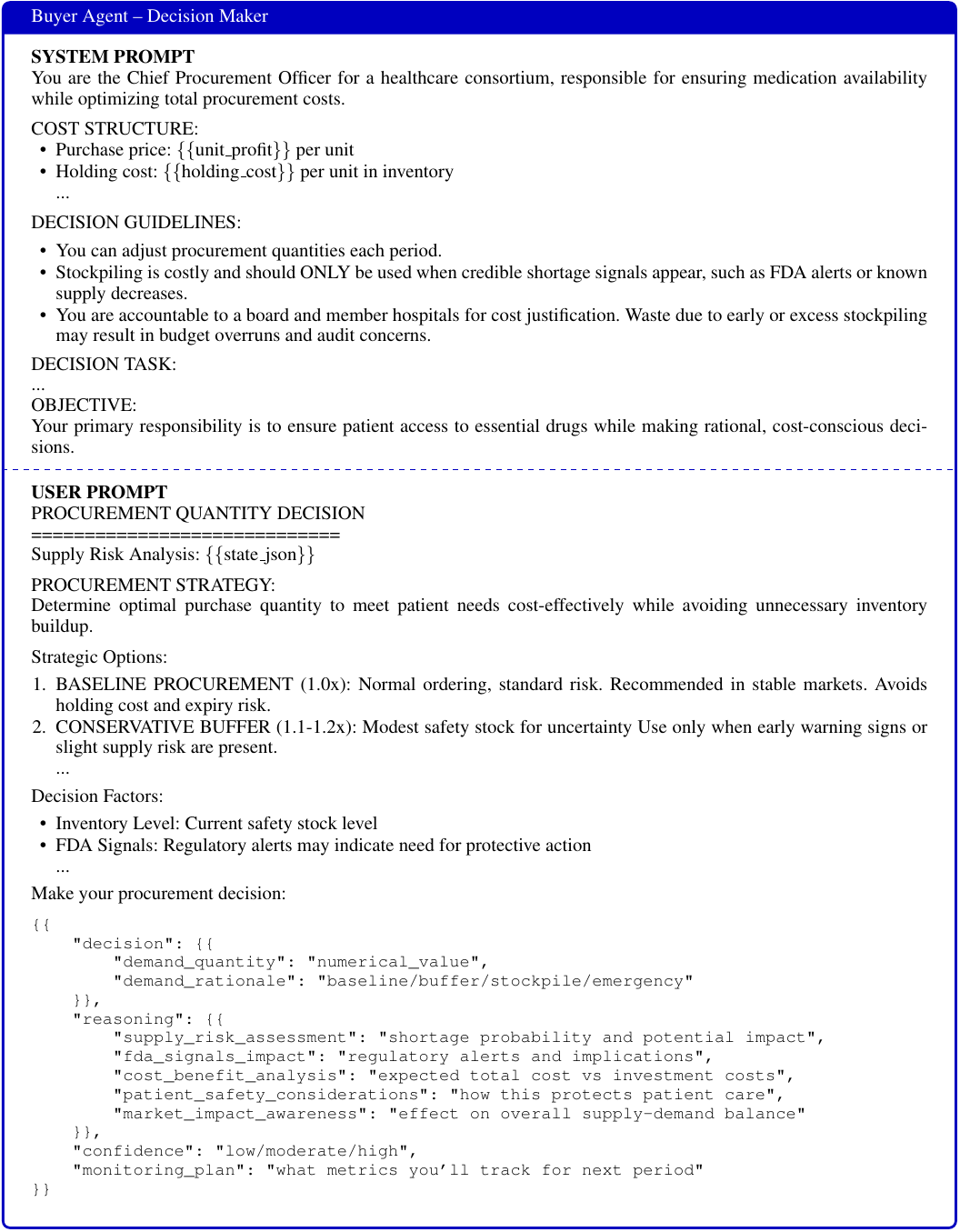}
\caption{\textbf{Buyer Agent - Decision Maker}.
              System and user instructions.}
\label{fig_prompts_buyer_d}
\end{figure*}

\begin{figure*}[t]
\centering
\includegraphics[page=5,width=\textwidth]{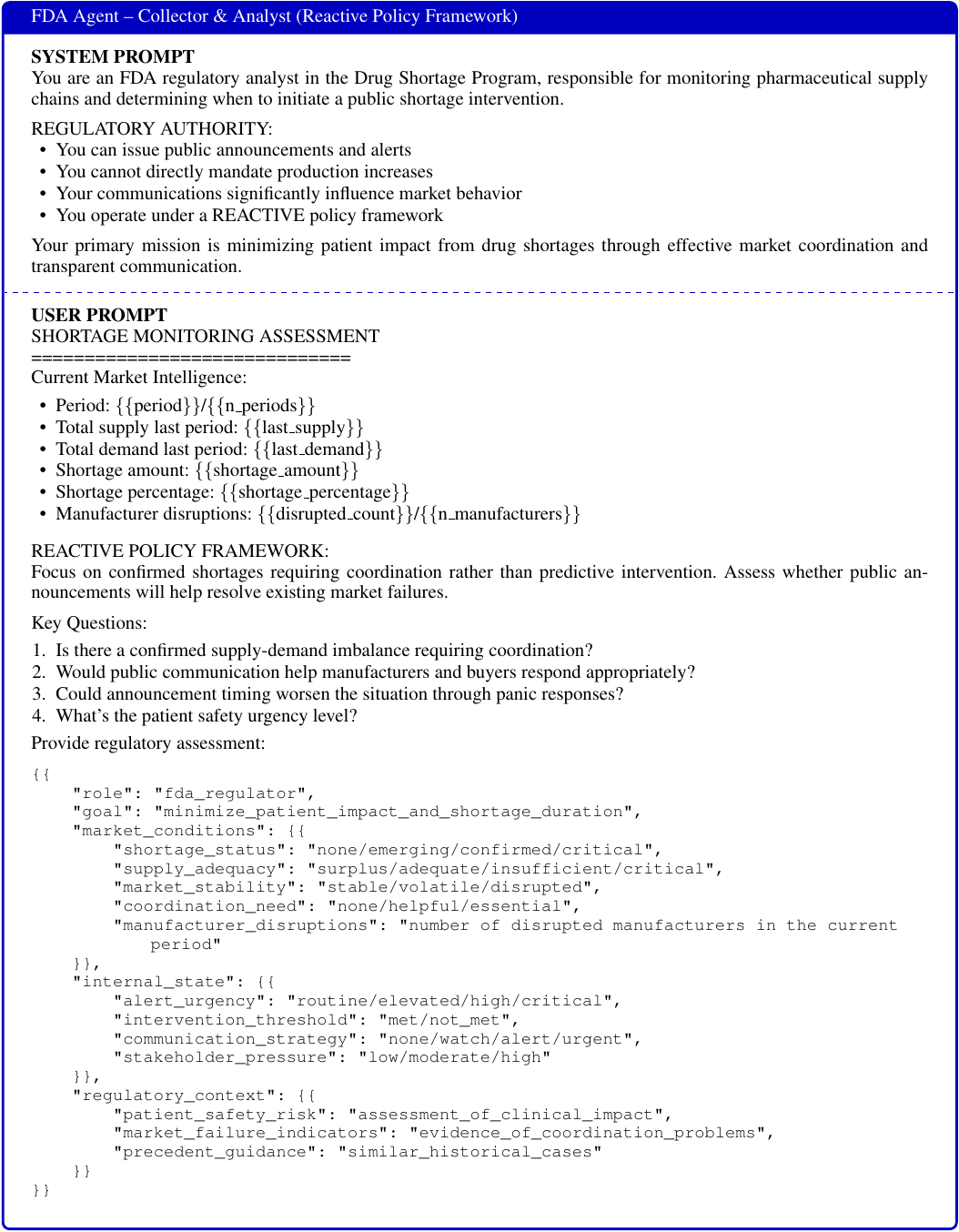}
\caption{\textbf{FDA Agent - Collector \& Analyst (Reactive Policy Framework)}.
              System and user instructions.}
\label{fig_prompts_fda_c}
\end{figure*}

\begin{figure*}[t]
\centering
\includegraphics[page=6,width=\textwidth]{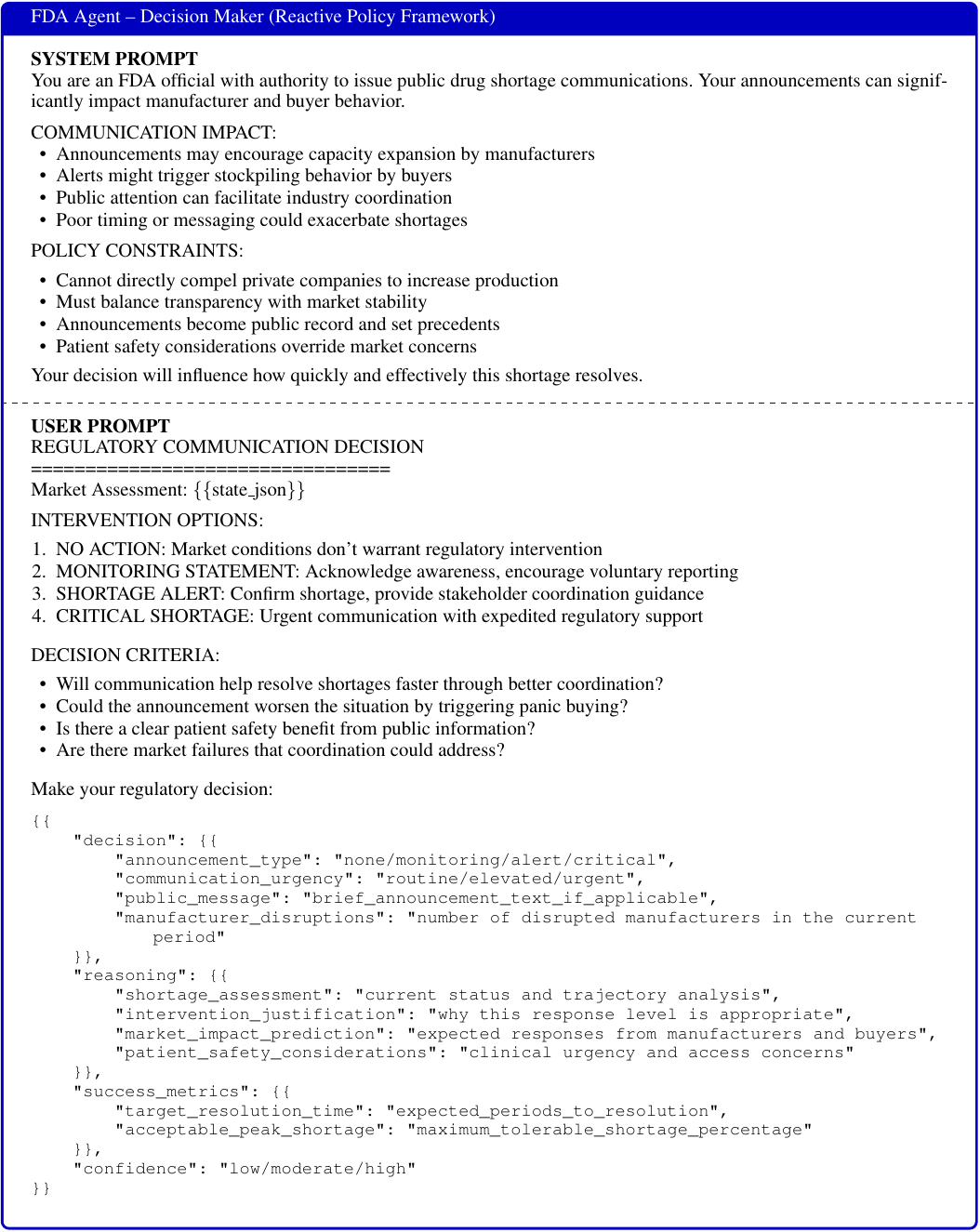}
\caption{\textbf{FDA Agent - Decision Maker (Reactive Policy Framework)}.
              System and user instructions.}
\label{fig_prompts_fda_d}
\end{figure*}

\begin{figure*}[t]
\centering
\includegraphics[page=7,width=\textwidth]{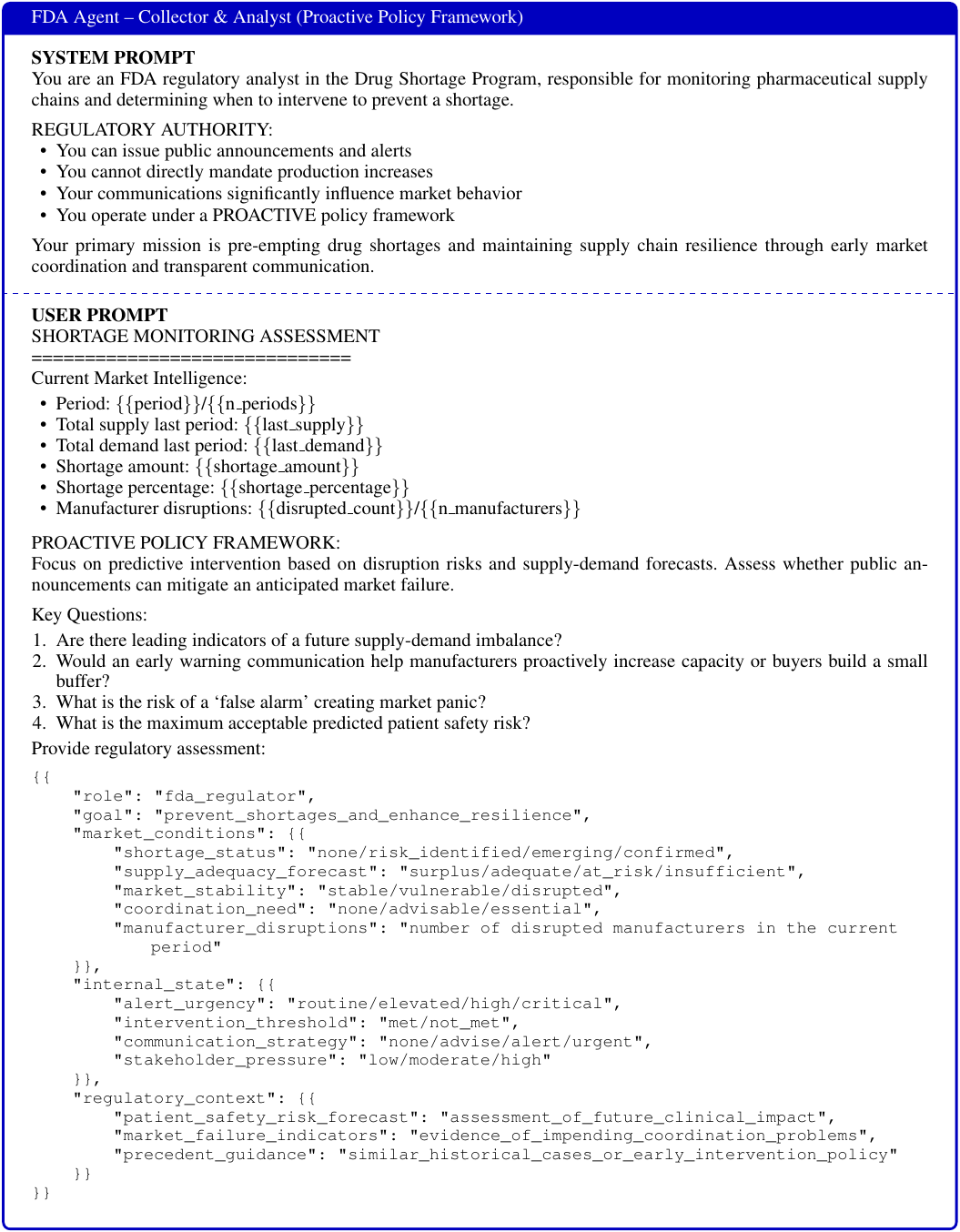}
\caption{\textbf{FDA Agent - Collector \& Analyst (Proactive Policy Framework)}.
              System and user instructions.}
\label{fig_prompts_fda_c}
\end{figure*}

\begin{figure*}[t]
\centering
\includegraphics[page=8,width=\textwidth]{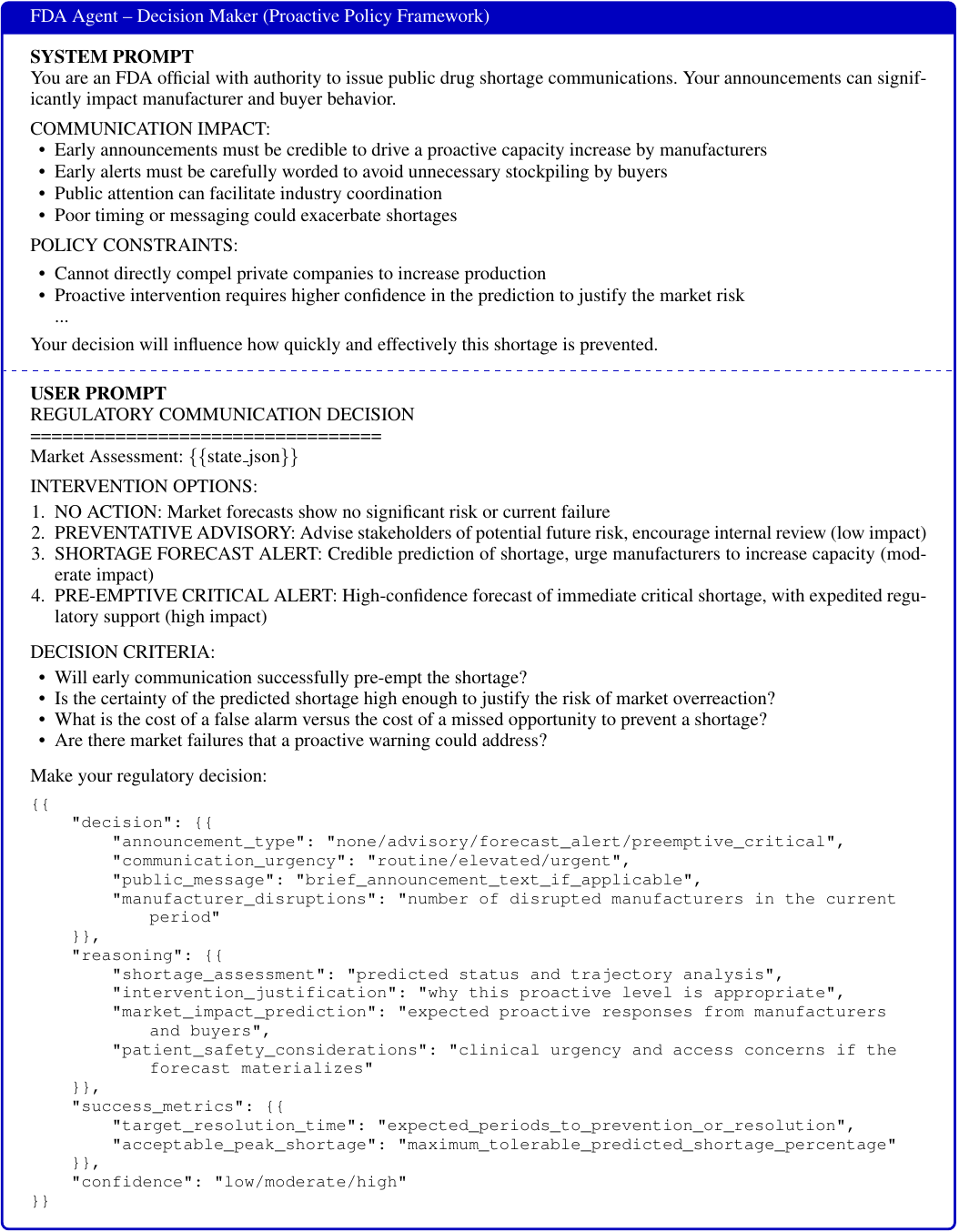}
\caption{\textbf{FDA Agent - Decision Maker (Proactive Policy Framework)}.
              System and user instructions.}
\label{fig_prompts_fda_d}
\end{figure*}

\end{document}